\documentclass[twocolumn,showpacs,showkeys,preprintnumbers,amsmath,amssymb,superscriptaddress]{revtex4}

\usepackage{graphics}
\usepackage{graphicx}
\usepackage{dcolumn}
\usepackage{bm}

\usepackage{amssymb}

\usepackage{enumerate}

\begin{document} 


\title{Proton-air cross section measurement with the ARGO-YBJ cosmic ray experiment} 

\author{G. Aielli} 
\affiliation{Dipartimento di Fisica dell'Universit\`a ``Tor Vergata'' - via della Ricerca Scientifica 1 - 00133 Roma - Italy}  
\affiliation{Istituto Nazionale di Fisica Nucleare - Sezione di Tor Vergata - via della Ricerca Scientifica 1 - 00133 Roma - Italy}  
 
\author{C. Bacci} 
\affiliation{Dipartimento di Fisica dell'Universit\`a ``Roma Tre'' - via della Vasca Navale 84 - 00146 Roma - Italy}  
\affiliation{Istituto Nazionale di Fisica Nucleare - Sezione di Roma3 - via della Vasca Navale 84 - 00146 Roma - Italy}  
 
\author{B. Bartoli} 
\affiliation{Istituto Nazionale di Fisica Nucleare - Sezione di Napoli - Complesso Universitario di Monte Sant'Angelo - via Cintia - 80126 Napoli - Italy}  
\affiliation{Dipartimento di Fisica dell'Universit\`a di Napoli - Complesso Universitario di Monte Sant'Angelo - via Cintia - 80126 Napoli - Italy}  
 
\author{P. Bernardini} 
\affiliation{Dipartimento di Fisica dell'Universit\`a del Salento - via per Arnesano - 73100 Lecce - Italy}  
\affiliation{Istituto Nazionale di Fisica Nucleare - Sezione di Lecce - via per Arnesano - 73100 Lecce - Italy}  
 
\author{X.J. Bi} 
\affiliation{Key Laboratory of Particle Astrophyics - Institute of High Energy Physics - Chinese Academy of Science - P.O. Box 918 - 100049 Beijing - P.R. China}  
 
\author{C. Bleve}  
\affiliation{Dipartimento di Fisica dell'Universit\`a del Salento - via per Arnesano - 73100 Lecce - Italy}  
\affiliation{Istituto Nazionale di Fisica Nucleare - Sezione di Lecce - via per Arnesano - 73100 Lecce - Italy}  
 
\author{P. Branchini}  
\affiliation{Istituto Nazionale di Fisica Nucleare - Sezione di Roma3 - via della Vasca Navale 84 - 00146 Roma - Italy}  
 
\author{A. Budano}  
\affiliation{Istituto Nazionale di Fisica Nucleare - Sezione di Roma3 - via della Vasca Navale 84 - 00146 Roma - Italy}   
 
\author{S. Bussino} 
\affiliation{Dipartimento di Fisica dell'Universit\`a ``Roma Tre'' - via della Vasca Navale 84 - 00146 Roma - Italy}  
\affiliation{Istituto Nazionale di Fisica Nucleare - Sezione di Roma3 - via della Vasca Navale 84 - 00146 Roma - Italy}  
 
\author{A.K. Calabrese Melcarne} 
\affiliation{Istituto Nazionale di Fisica Nucleare - CNAF - viale Berti-Pichat 6/2 - 40127 Bologna - Italy} 
 
% ...


\collaboration{The ARGO-YBJ Collaboration} 
\noaffiliation


\date{\today}

\begin{abstract} 
The proton-air cross section in the energy range 1-100$\,$TeV has been measured by the ARGO-YBJ 
cosmic ray experiment.
The analysis is based on the flux attenuation for different atmospheric depths (i.e. zenith angles) 
and exploits the detector capabilities of selecting the shower development stage by means of
hit multiplicity, density and lateral profile measurements at ground. 
The effects of shower fluctuations, the contribution of heavier primaries and the uncertainties of the hadronic
interaction models, have been taken into account.
The results have been used to estimate the total proton-proton cross section at center of mass energies
between 70 and 500$\,$GeV, where no accelerator data are currently available.  
\end{abstract}



\maketitle

\section{Introduction} 
\label{sec:intro}

In the last decades, high-energy proton-antiproton colliders allowed extending the study of hadronic interactions 
up to center of mass energy $\sqrt{s} = 1.8 \,$TeV \cite{geich-gimbel1989,avila2002,abe1994}.
Moreover LHC experiments at CERN are expected to analyze proton-proton collisions 
and to measure total cross sections at $\sqrt{s} = 14 \,$TeV \cite{osterberg2008}. 
Anyway, at energies exceeding $\sqrt{s} \sim 70\,$GeV, where total cross sections start increasing, 
the knowledge of hadronic interactions is limited by experimental uncertainties 
and/or lack of data \cite{pdg2008,block2006}.  

The study of cosmic ray (CR) induced showers provides a unique opportunity to explore hadron 
interactions in an energy range which not only covers the LHC region, 
but also extends well beyond it, the natural beam of primary particles spanning up 
to extreme values \cite{abraham2008}.
Furthermore, some phase space regions cannot be easily accessed at colliders, where, as an example, 
most of the produced particles might escape undetected in the beam pipe in the case of {\it soft}, 
i.e. low momentum transfer, hadronic interactions.
On the contrary, {\it soft } collisions play a major role in the development of CR-induced showers, 
carrying energy deep down in the atmosphere. 
In currently used interaction models, these collisions are treated by means of the Gribov-Regge Theory (GRT) \cite{gribov1983} 
via {\it Pomerons} exchange \cite{werner1993}, while a perturbative QCD approach can be appropriate 
at most collider experiments. Finally, it must be considered that, in the case of CR experiments, the
collision targets are light nuclei (i.e. N, O, etc., hereafter {\it ``air''} nuclei)
and nuclear effects cannot be neglected. 
These aspects make somehow complementary the two approaches and cause their comparison to be extremely 
interesting from the point of view of very high energy (astro)particle physics.

In CR experiments, the proton-air (hereafter p-air) cross section can be measured in several 
ways depending on the energy range and detection technique.
Suitable models can then be used to estimate the p-p cross section \cite{block2006}.
Because of the rapidly falling CR energy spectrum, the use of ground based detectors is mandatory.
This obviously implies that the method systematics are larger with respect to accelerator experiments,
the interaction initial state being not under control. 
Moreover, as pointed out in \cite{engel1998}, these experiments 
are actually sensitive only to the so-called {\it production} cross section defined as
\begin{equation}
\label{eq:prodxsection}
  \sigma^{prod}_{p-air} = \sigma^{tot}_{p-air} - \sigma^{el}_{p-air} - \sigma^{qel}_{p-air}
\end{equation}
where $\sigma^{tot}_{p-air}$ and $\sigma^{el}_{p-air}$ are the total and the elastic cross sections respectively, 
while $\sigma^{qel}_{p-air}$ refers to quasi-elastic processes in which the nucleus gets excited without 
secondary particle production.
There is some ambiguity in the literature, since the quantity $\sigma^{prod}_{p-air}$, as defined in 
Eq.\ref{eq:prodxsection}, has also been referred to as {\it inelastic} or {\it absorption} cross section. 
Here we will follow the terminology introduced in \cite{engel1998}.
For the sake of simplicity, in the following we will refer to $\sigma^{prod}_{p-air}$
 simply as the {\it p-air cross section} or $\sigma_{p-air}$.

The p-air cross section is measured by evaluating the absorption of the primary 
proton flux penetrating the atmosphere. 
In principle, the distribution of the atmospheric depth, $X_0$, of the first interaction point
of CR protons could give a measurement of the interaction length $\lambda_{p-air}$ 
(and then the cross section $\sigma_{p-air}$) in a straightforward way.
Unfortunately, this information cannot be currently accessed and indirect methods have to be used.
Up to now three experimental approaches have been adopted, 
depending on the primary energy range and detection techniques. 
Their comprehensive description is out of the scope of this paper. 
Here we will shortly list them and we will then focus, in the following sections, 
on the technique adopted in this work.

The first method dates back to 1960s and actually estimates the flux of protons that reach the detection level
without having yet interacted producing an hadronic shower 
(see \cite{yodh1983,gaisser1987,mielke1994} and references therein).
This is made by measuring the flux of unaccompanied hadrons crossing high altitude calorimeters equipped with 
a suitable veto system. The flux is then compared with the measured primary cosmic ray flux at the top 
of the atmosphere.  One of the intrinsic systematic errors of this approach comes from the uncertainty on the 
primary proton flux that is used in the estimate of the p-air cross section. 
By using this technique, one of the first evidences of the rising of the total p-p cross section with 
energy has been established \cite{yodh1972}.
The reduction of the proton flux prevents the adoption of this method at high energies, where the 
experiments must rely on the observed extensive air showers (EAS) initiated by CR primaries at the top of 
the atmosphere.

EAS arrays can evaluate the primary proton absorption in the atmosphere by measuring the shower 
rates at different zenith angles (i.e. atmospheric depths) once the shower development stage and energy are 
constrained by observations at ground 
(see \cite{hara1983,honda1993,knurenko,aglietta2009} and references therein). 
The measured number of muons, $N_\mu$, is generally used to evaluate the shower energy
(and to select proton-initiated events), while the number of charged particles at the detection level, $N_e$,
gives an estimate of the shower development stage.  
In this case an important role can be played by the shower-to-shower fluctuations,
that might decrease the experimental capability of properly measuring the p-air cross section \cite{alvarez2002,alvarez2004}.
Since these effects depend on cosmic ray energy and mass composition, on detector 
capabilities and location, and on actual analysis procedures,
the sensitivity of each experiment has to be checked with detailed simulations.
This technique, traditionally known as the $N_e - N_{\mu}$ method, has a further source of systematic uncertainty which 
comes from the appreciable model dependence of the predicted muon numbers \cite{pierog2008,ulrich2009}.

At even larger energies, the shower longitudinal profile can be directly measured with the air fluorescence technique
(and/or using the lateral distribution of \v{C}erenkov light at ground).
The atmospheric depth of the maximum shower development stage, $X_{max}$, gives information on the CR beam absorption,
i.e. on the p-air cross section (see \cite{knurenko,baltrusaitis1984,belov2007} and references therein).
In this case, one of the largest systematics comes from the uncertanties on CR composition.
This effect is also present in the previously mentioned approaches, but it becomes more and more important with increasing 
primary energy. In particular, the effect of CR primaries heavier than protons is negligible in the first approach,
while it is small in the second one and a correction can be applied on the basis of CR composition measurements.
At extremely high energies the CR composition is not sufficiently well known in order for that correction to be applied
without introducing a sizeable systematic uncertainty.

Once the p-air cross section is measured, theoretical models can be used to give the total p-p cross section.
This conversion is essentially based on the Glauber theory \cite{glauber1970} and has been discussed in several works
\cite{block2006,engel1998,gaisser1987,durand1990,kopeliovich1989,bellandi1995,wibig1998,block2007}.

In this paper we will report on the measurement of the {\it production} cross section between 1-100$\,$TeV CR protons 
and air nuclei with the ARGO-YBJ experiment (see Sec.\ref{sec:argo}).
The analysis is based on the flux attenuation for different atmospheric depths, 
i.e. zenith angles, and exploits the high detector accuracy in reconstructing the shower properties at ground.
This allows a different approach with respect to the standard $N_e -N_\mu$ method 
used in EAS experiments that operate at larger energies and with lower shower front sampling capabilities.
The results have also been used to estimate the total p-p cross section at center of mass energies
between 70 and 500$\,$GeV, where no accelerator data are currently available \cite{pdg2008}.
The adopted strategy is discussed in  Sec.\ref{sec:meatechnique}, while details on data selection
and analysis procedure are given in Sec.\ref{sec:analysis}.
Results are then given in Sec.\ref{sec:discussion} together with the discussion of the method systematics. 
Conclusions are reported in Sec.\ref{sec:conclus}.

\begin{figure*}
 \begin{center}
    \includegraphics [width=0.9\textwidth]{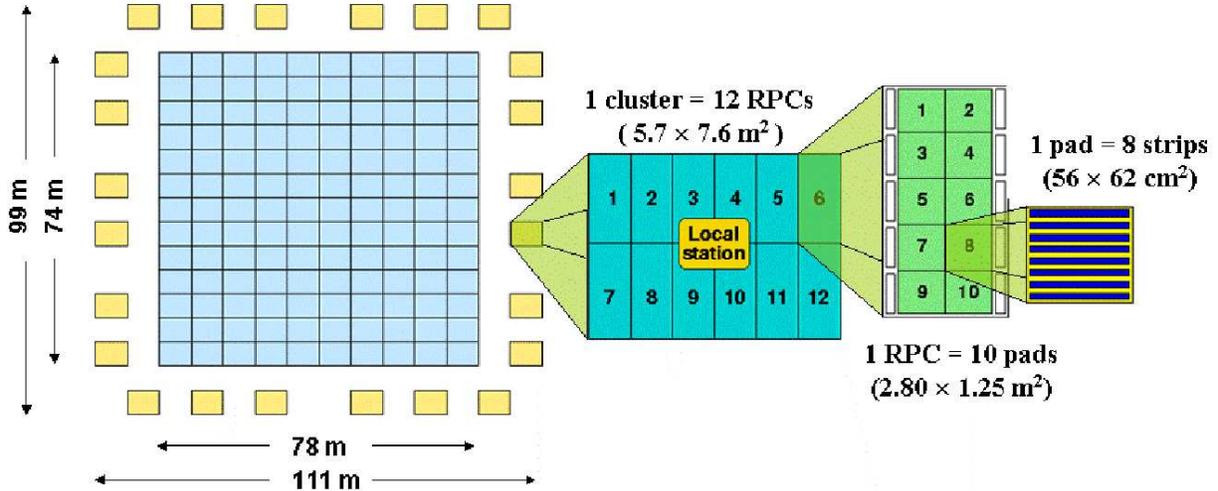}
 \caption{The ARGO-YBJ detector setup. The {\it cluster} (12 RPCs) is the basic detector and DAQ unit
 equipped with a {\it Local Station} for its read-out.
 The full-coverage central carpet is made by 130 clusters, the guard-ring by 24 clusters. The pad
 is the timing pixel and is further divided into 8 strips (see text).}
\label{fig:setup}
 \end{center}
\end{figure*}

\begin{figure*}
 \begin{center}
   \begin{tabular}{cc}
    \includegraphics [width=.42\textwidth]{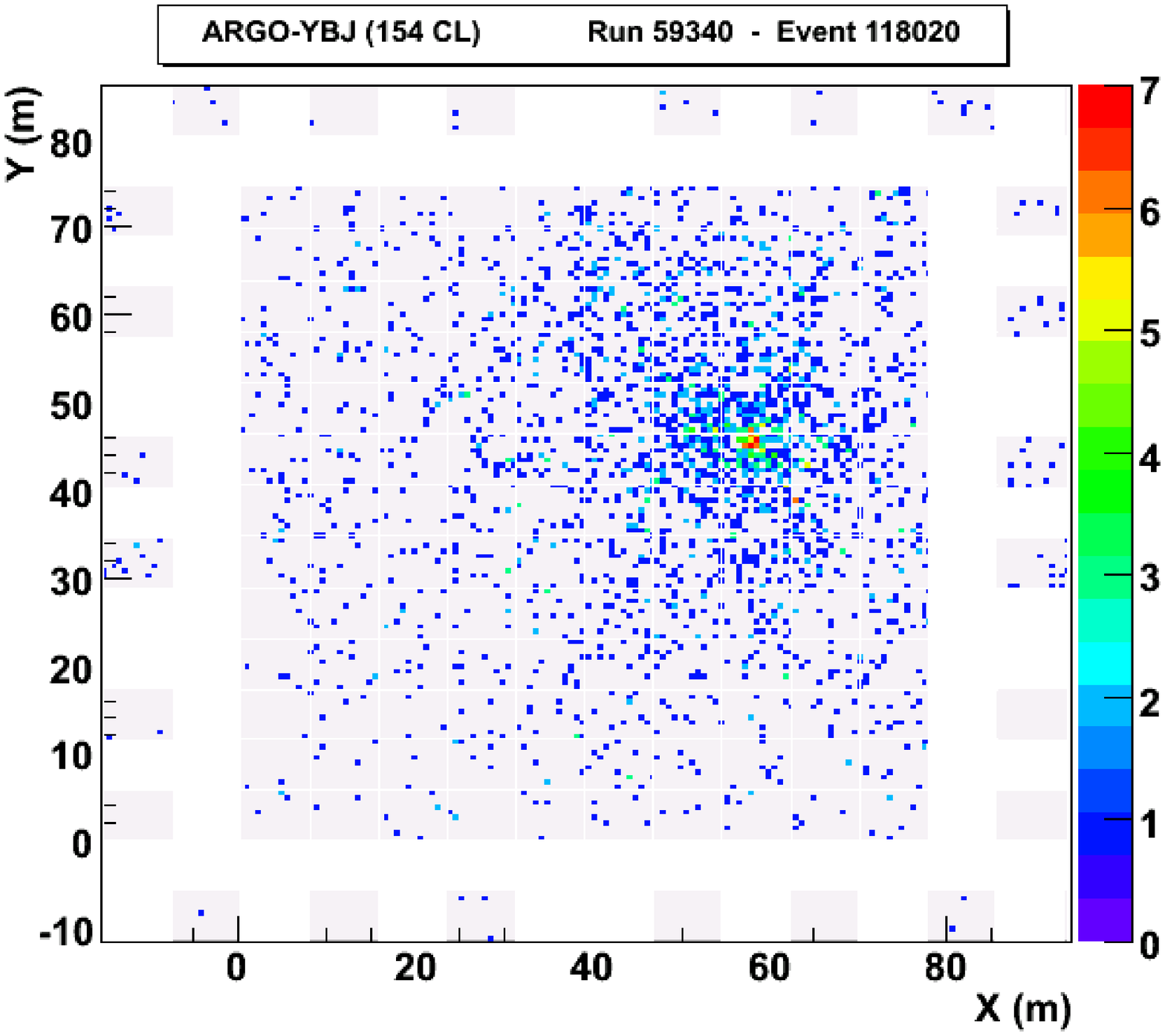} &
    \includegraphics [width=.42\textwidth]{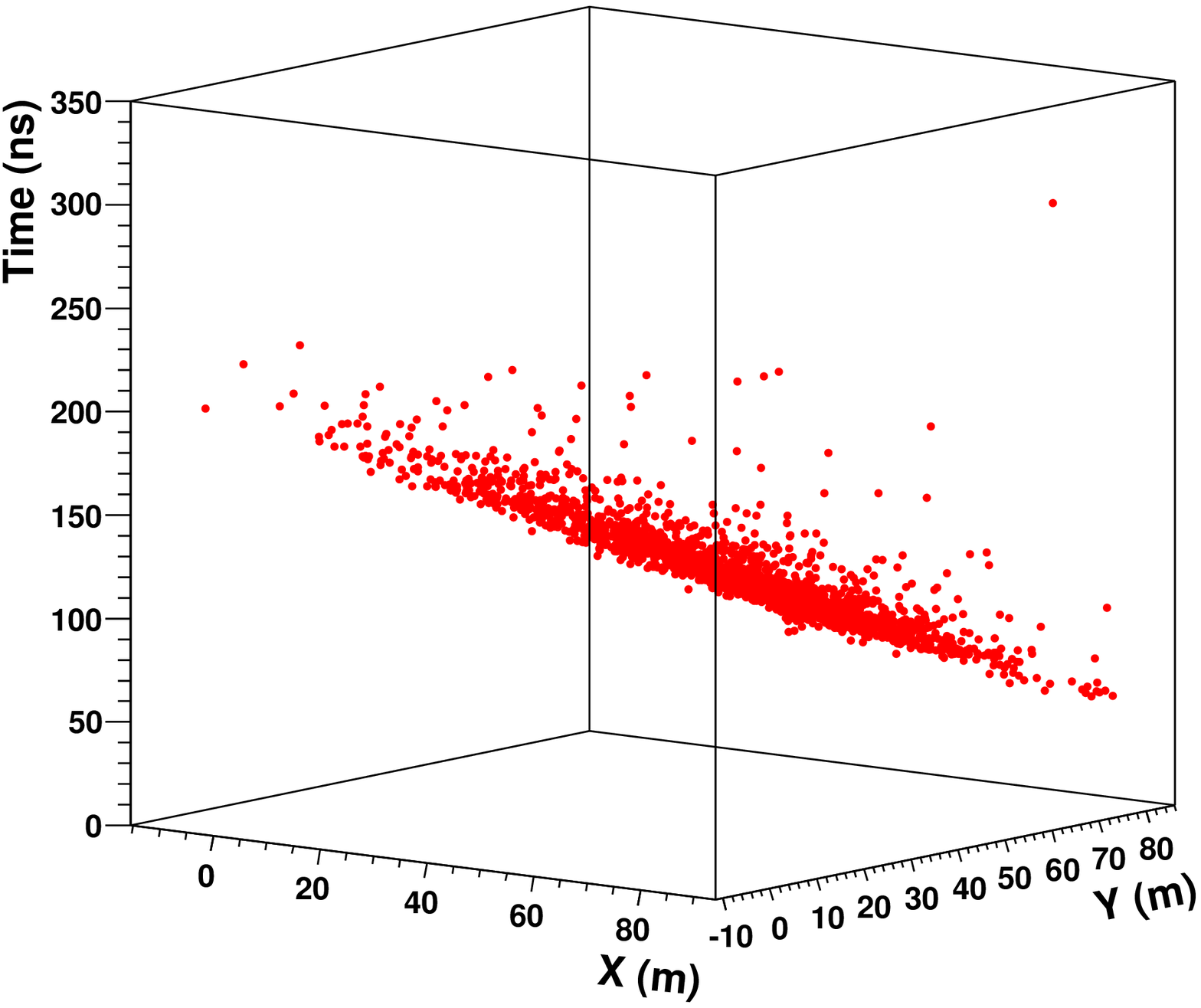}   \\
   \end{tabular}
 \caption{Two different views of a detected shower. The hit map at ground is given on the left, the color code
          representing the strip multiplicity of each fired pad. A space-time view of the detected shower front 
          is given on the right.}
 \label{fig:event}
 \end{center}
\end{figure*}

\section{The ARGO-YBJ experiment} 
\label{sec:argo} 

Cosmic ray physics in the $1-1000\,$TeV energy range and very high energy (VHE) $\gamma$-ray astronomy are the 
main scientific goals of the ARGO-YBJ experiment, a full coverage extensive air shower array 
resulting from a collaboration between Chinese and Italian institutions \cite{bacci2002,demitri2007cris}.
The detector is operating in the YBJ International Cosmic Ray Observatory, located in the village of Yangbajing, 
about 90$\,$km north-west from Lhasa, in the Tibet region (People's Republic of China) at an altitude of 4300$\,$m above sea level, 
corresponding to a vertical atmospheric depth of about $610 \,$g/cm$^2$.
The apparatus, extensively described in \cite{bacci2002,demitri2007cris}, 
is a single layer detector logically divided into 154 units called {\it clusters} ($7.64 \times 5.72 \,$m$^2$)
each made by 12 Resistive Plate Chambers (RPCs) operated in streamer mode (see Fig.\ref{fig:setup}).
Each RPC ($1.25 \times 2.80 \,$m$^2$) is read out by using 10 pads ($62 \times 56 \,$cm$^2$), which are  
further divided into 8 strips ($62 \times 7 \,$cm$^2$) providing a larger particle counting 
dynamic range \cite{aielli2006}. 
The signals coming from all the strips of a given pad are sent to the same channel of a multihit TDC.  
The whole system, in smooth data taking from mid 2006, provides a single hit (pad) time resolution at the level of 1 ns,
which allows a complete and detailed three-dimensional reconstruction of the shower front with 
unprecedented space-time resolution (see Fig.\ref{fig:event}).
In order to fully investigate transient phoenomena with an even smaller energy threshold, 
an independent {\it scaler mode} acquisition system has also been put in operation \cite{aielli2008}.
Finally, a system for the RPC analog charge readout \cite{creti2005} from larger pads, 
each one covering half a chamber (the so called {\it big pads}), is now being implemented.
This will allow extending the detector operating range from about 100$\,$TeV up to PeV energies.

A suitable calibration procedure has been developed in order to remove systematic time offsets 
in the read-out chain (due to front-end boards, TDC boards, cable lengths, etc.) among the 18480 
different channels \cite{aielli2009}. Many checks have been made in order to ensure the 
correctness of the procedure. One of these is made through the analysis of the position, size and shape
of the reconstructed Moon and Sun shadows in the cosmic ray flux \cite{dali2008,disciascio2008}.

Due to the huge amount of data, event reconstruction is performed by using GRID-based object oriented technologies 
\cite{stanescu2007}.
In the first event reconstruction step, the detected shower front is fitted to a plane.
Actually, detailed simulations show that particles within several tens of meters from the shower core 
do form a curved front, which can be well approximated with a conical shape \cite{calabrese2007}.
This is confirmed by experimental data and is suitably exploited in the second step of the event reconstruction.
A reliable identification of the shower core position up to the edge, and slightly beyond, the active carpet 
is made through a Maximum Likelihood-based algorithm, using a NKG-like \cite{nkg} lateral distribution
function \cite{disciascio2003}. 
Space-time coordinates (positions and times of fired pads) are then 
fitted to a cone whose axis crosses the core position at ground \cite{martello2003}.
As a result a very good angular resolution, $\sigma_{ang}$ is obtained \cite{disciascio2008}. 
Obviously the actual value depends on the CR primary energy and mass ($\gamma$, p, He, ...), 
and on the specific event selection criteria adopted in a given analysis.
In our case, we get $\sigma_{ang}(E\sim 5\,$TeV)$\, \simeq 0.3^\circ$ and 
$\sigma_{ang}(E \gtrsim 10\,$TeV)$\, \lesssim 0.2^\circ$, for primary protons.
This is particularly important, since it implies that no significant systematics are introduced 
by the detector angular resolution in the zenith angle distributions used for 
the cross section measurement (see Sec.\ref{sec:discussion}).

\section{The analysis strategy}
\label{sec:meatechnique}

\begin{figure*}
 \begin{center}
  \begin{tabular}{cc}
    \includegraphics [width=.42\textwidth]{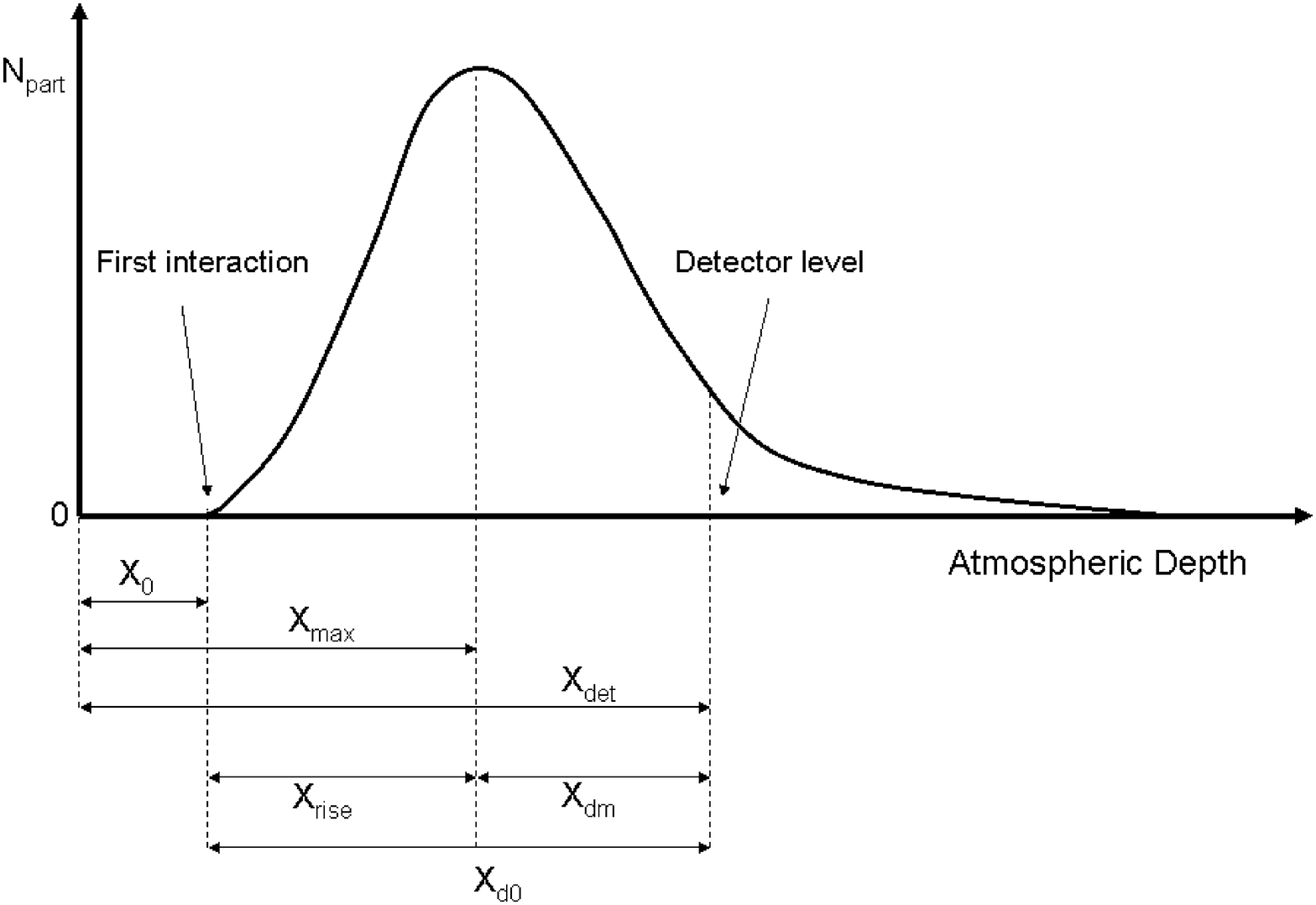} &
    \includegraphics [width=.42\textwidth]{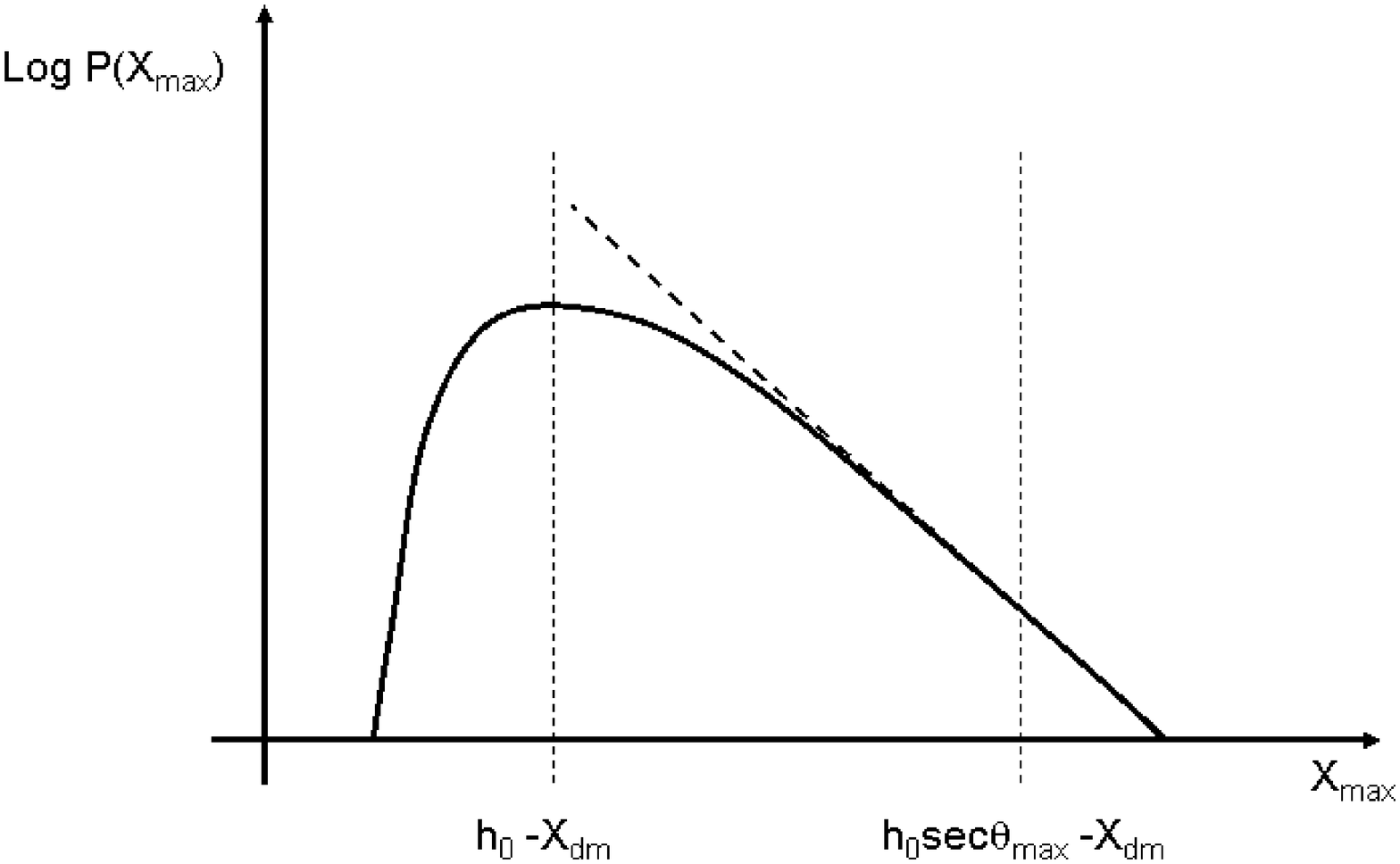}  
  \end{tabular}
 \caption{Simplified scheme of the shower longitudinal development in the atmosphere (left panel)
          and of the $X_{max}$ distribution (right panel), together with the definitions 
          of the quantities used in the text}
\label{fig:longdev}
 \end{center}
\end{figure*}

For a primary energy interval and for a given distance $X_{dm}$ between the detector and the shower maximum 
(see Fig.\ref{fig:longdev}), the frequency of showers as a function of zenith angle $\theta$ 
is directly related to the probability distribution of the depth of the shower maximum
$P(X_{max})$, where $X_{max} = h_0 sec\theta - X_{dm}$ and $h_0$ is the observation vertical depth.
The shape of $P(X_{max}$=$X_{0}$+$X_{rise})$ is given by the folding of the exponential dependence 
of the depth of the first interaction point $X_0$ 
(i.e. $e^{-X_0 / \lambda_{p-air}}$ with $\lambda_{p-air} (g/cm^2) \simeq 2.41 \times 10^4 / \sigma_{p-air}(mb)$), 
with the probability distribution of $X_{rise}$, which includes
the fluctuations of the shower development up to its maximum. 
Then the effects of the limited experimental resolution have to be also taken into account.
For sufficiently large $X_{max}$ values, $P(X_{max})$ tends to be  
a simple exponential with a characteristic length $\Lambda = k \lambda_{p-air}$, where 
$k$ depends on hadronic interactions and on the shower development in the atmosphere and its 
fluctuations \cite{block2006,ulrich2009,ellsworth1982,pryke2001}.
The actual value of $k$ must be evaluated with a full Monte Carlo (MC) simulation and it depends on
the experimental approach, the primary energy and the detector response. 
It might also depend on the features of the hadronic interaction model adopted in the simulation, 
this being a further source of systematic uncertainty. Anyway several studies showed that this dependence 
is generally small \cite{block2006,aglietta2009,pryke2001}.
Finally, the contribution of cosmic rays heavier than protons has to be estimated  and minimized in 
order to get an unbiased proton-air cross section measurement.

As discussed in Sec.\ref{sec:intro}, experiments using the air fluorescence technique have direct access to $P(X_{max})$, 
while EAS detectors measuring the particles at ground might sample it through the flux dependence on zenith angle,
once $X_{dm}$ (or the shower age) has been fixed or constrained, within the limits of detector capabilities.
In these cases the combination of CR energy, detector vertical depth, angular and  $X_{dm}$ ranges, 
actually define the part of the $X_{max}$ distribution that can be accessed. 
A good performance is obtained if the exponential tail of $P(X_{max})$ can be sampled.
As it is schematically shown in Fig.\ref{fig:longdev}, this requires selecting showers with the maximum development
not far from the detection level (i.e. minimizing $X_{dm}$) and, obviously, exploring a zenith angle region as wide 
as possible.

If this is not the case, a flattening of the distribution might be observed, also due to shower fluctuations, 
resulting in a larger value of the parameter $k$ and a lower sensitivity to $\sigma_{p-air}$.
The measurement can be spoiled also by a very poor energy resolution. 
In this case both shower-to-shower fluctuations and the $X_{rise}$ dependence 
on the primary energy might contribute to a worsening of the analysis performance.
In some particular cases even the exponential behaviour can be lost \cite{ulrich2009}.

The ARGO-YBJ detector features and location (full coverage, angular resolution, fine granularity, 
small atmospheric depth, etc.), which ensure the capability of reconstructing the detected showers in a 
very detailed way, have been used to define the energy ranges and to constrain the shower ages.
In particular, different hit (i.e. strip) multiplicity windows have been used to select showers 
corresponding to different primary energies, while the information on particle density, lateral profile 
and shower front extension has been used to constrain $X_{dm}$ in the proper range.
These features allowed the observation of the exponential falling of shower frequencies, through the $sec\theta$ distribution

The fit to the angular distribution $R(\theta)$ gives the slope value $\alpha$, connected 
to the observed characteristic length $\Lambda$ through the relation $\alpha = h_0 / \Lambda$, being
\begin{equation}
  R(\theta) = A(\theta) \, R_0  \, e^{-\alpha (sec\theta - 1)} 
  \label{eq:sectheta}
\end{equation}
where $R_0~$=$~R(\theta$=$0)$, 
and the factor $A(\theta)$ takes into account the geometrical acceptance of each angular bin.

The same analysis chain is then applied to the simulated data sample (see next section). 
For each strip multiplicity (i.e. primary energy) interval, the fit to the $sec\theta$ distribution with the 
function of Eq.\ref{eq:sectheta} gives the value of $\Lambda^{MC}$.
The value of $k$, referring to each multiplicity bin, can then be evaluated as $k = \Lambda^{MC} / \lambda^{MC}_{p-air}$, 
where $\lambda^{MC}_{p-air}$ is known, in the corresponding energy region, from the adopted interaction model.

The experimental interaction length is obtained by correcting the observed characteristic length 
$\Lambda^{exp}_{CR-air}$ by the factor $k$ determined on the basis of the MC simulation:
$\lambda^{exp}_{CR-air} = \Lambda^{exp}_{CR-air} /  k$. 
Such value will give the measured p-air interaction length ($\lambda^{exp}_{p-air}$), once the effects of heavier
nuclei present in the primary cosmic ray flux have been taken into account. This has been made by evaluating the
change of the slope $\alpha$ (see Eq.\ref{eq:sectheta}) produced by the addition of the corresponding helium fraction 
to the proton primary flux in the MC simulation, the contribution of heavier nuclei being negligible.

The proton-air {\it production} cross section is then obtained from the previously mentioned relation:
$\sigma_{p-air}\,$(mb)$ = 2.41 \times 10^{4} / \lambda^{exp}_{p-air}\,$(g/cm$^2$), while several 
theoretical models can be used to get the corresponding total proton-proton cross section $\sigma_{p-p}$
(see Sec.\ref{sec:discussion}).

\section{Data Selection and Monte Carlo simulation}
\label{sec:analysis}

The analysis was applied to a data sample of about $6.5 \times 10^8$ events collected by the central part of the detector,
i.e. the $130$ adjoining clusters fully covering a surface of $78 \times 74 \,$m$^2$ (see Fig.\ref{fig:setup}). 
An inclusive trigger requiring at least 20 fired pads in a $420\,$ns time window was used.
In order to have both a small contamination of {\it external} events
(i.e. those events with the true core position outside the carpet but reconstructed inside)
and an angular resolution better than 0.5$^\circ$, only events with at least 500 fired strips were considered.
Moreover, the analysis was restricted to events with reconstructed zenith angle 
$\theta \le 40^\circ$. This was made in order to avoid effects due to the 
possible zenith angle dependence of the analysis cuts, that might occur at $\theta \gtrsim 45^\circ$
(see below).

\begin{figure*}
 \begin{center}
  \begin{tabular}{cc}
\includegraphics [width=0.4\textwidth,height=0.3\textwidth]{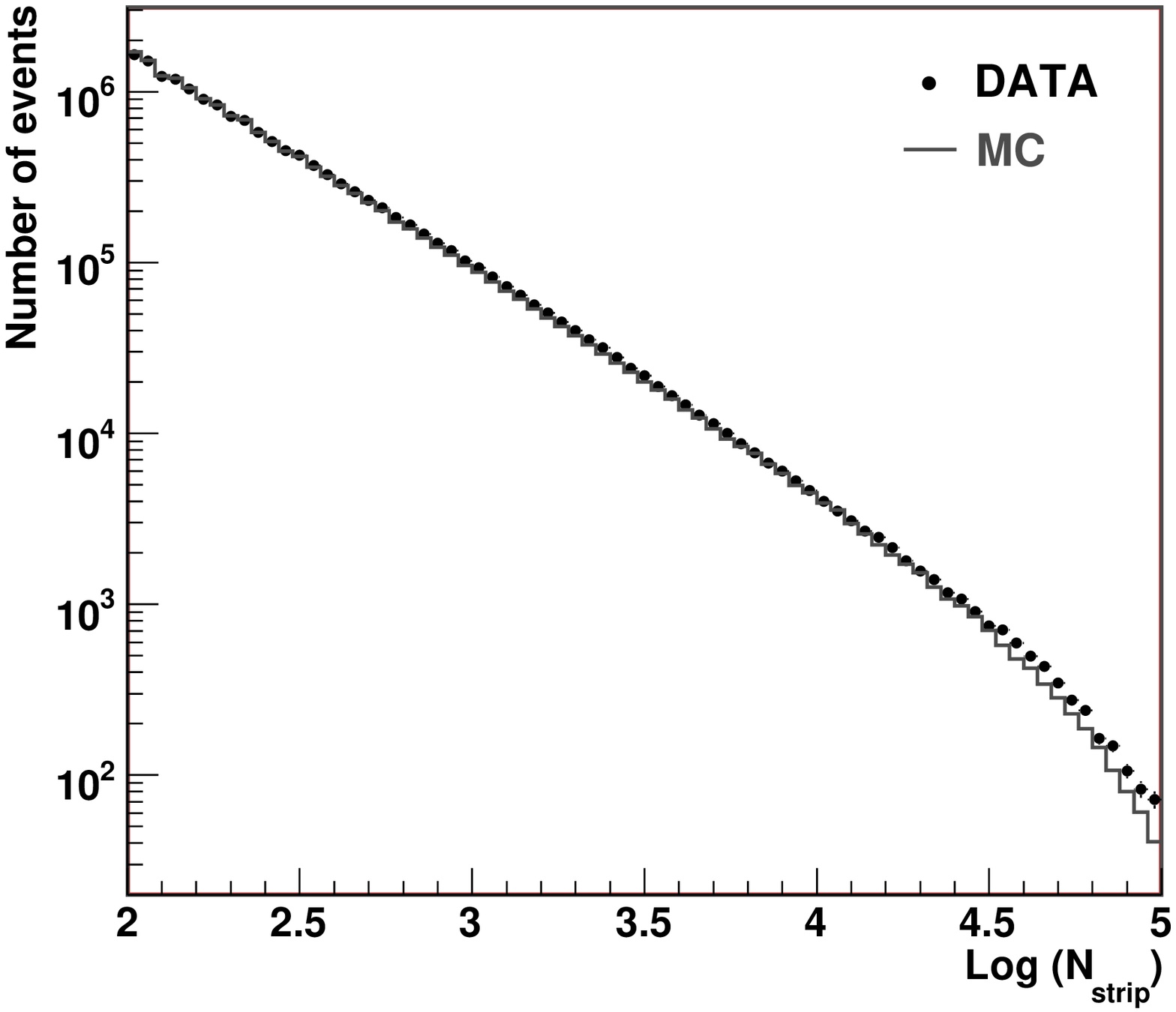} & 
\includegraphics [width=0.4\textwidth,height=0.3\textwidth]{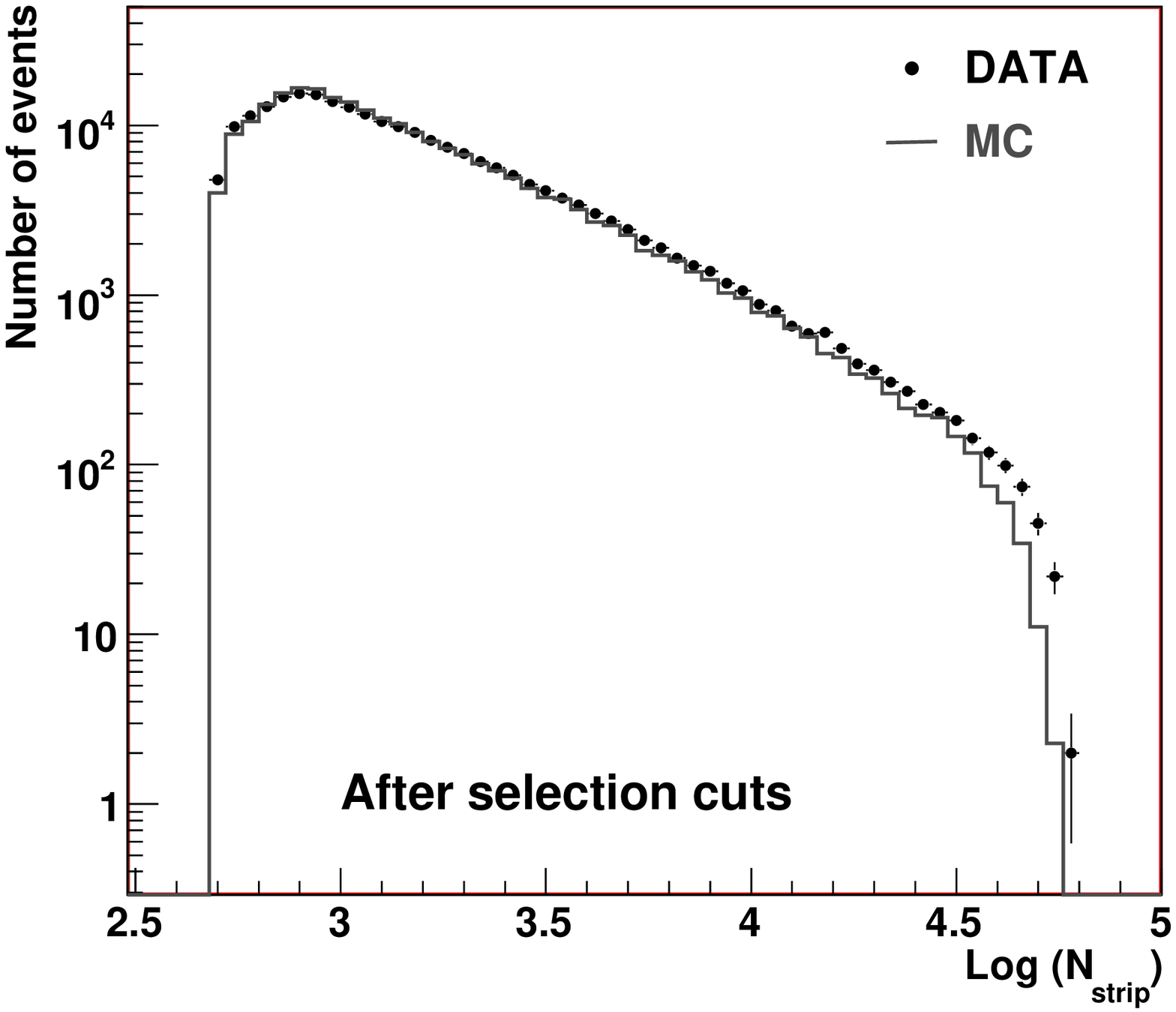} \\
\includegraphics [width=0.4\textwidth,height=0.3\textwidth]{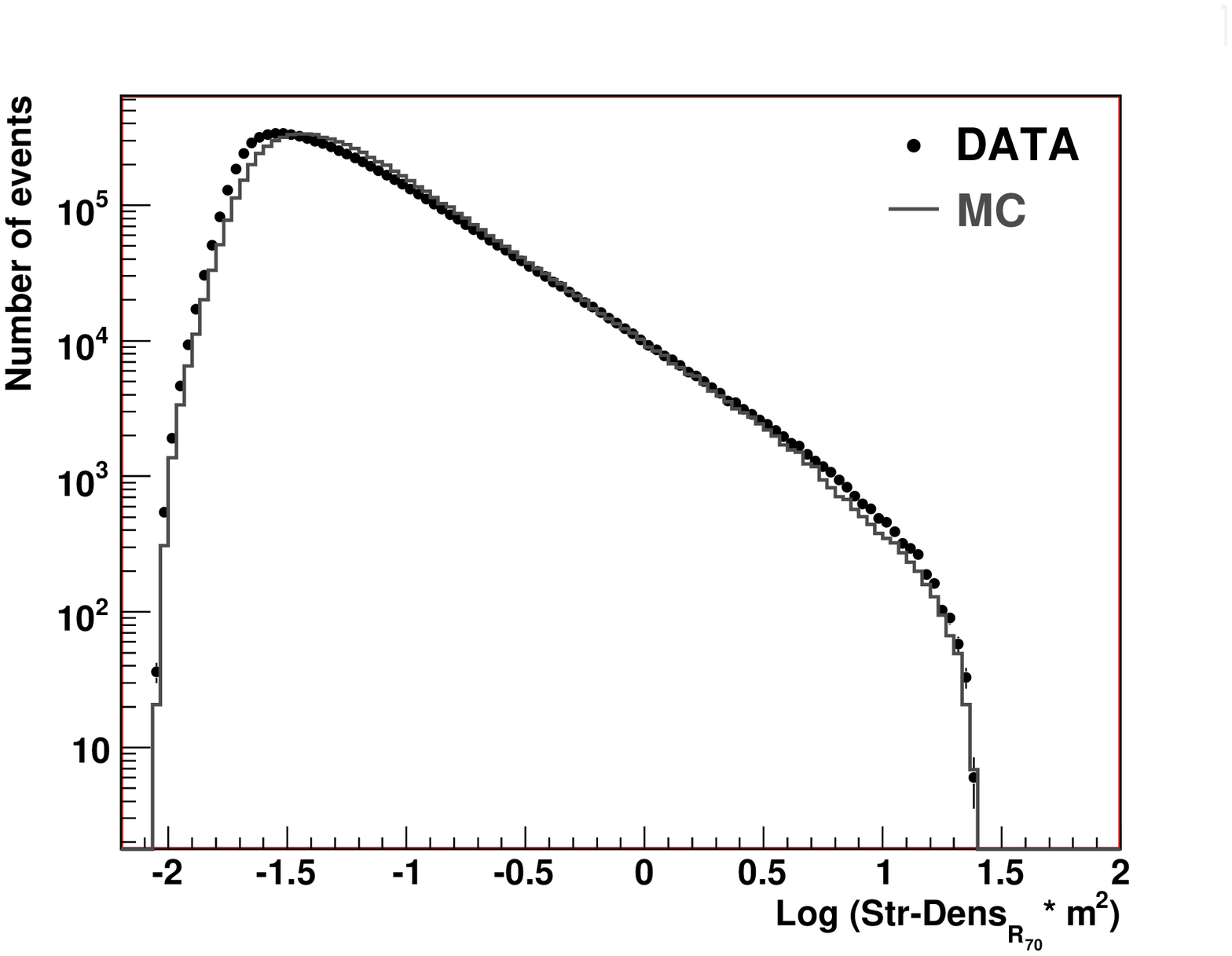} & 
\includegraphics [width=0.4\textwidth,height=0.3\textwidth]{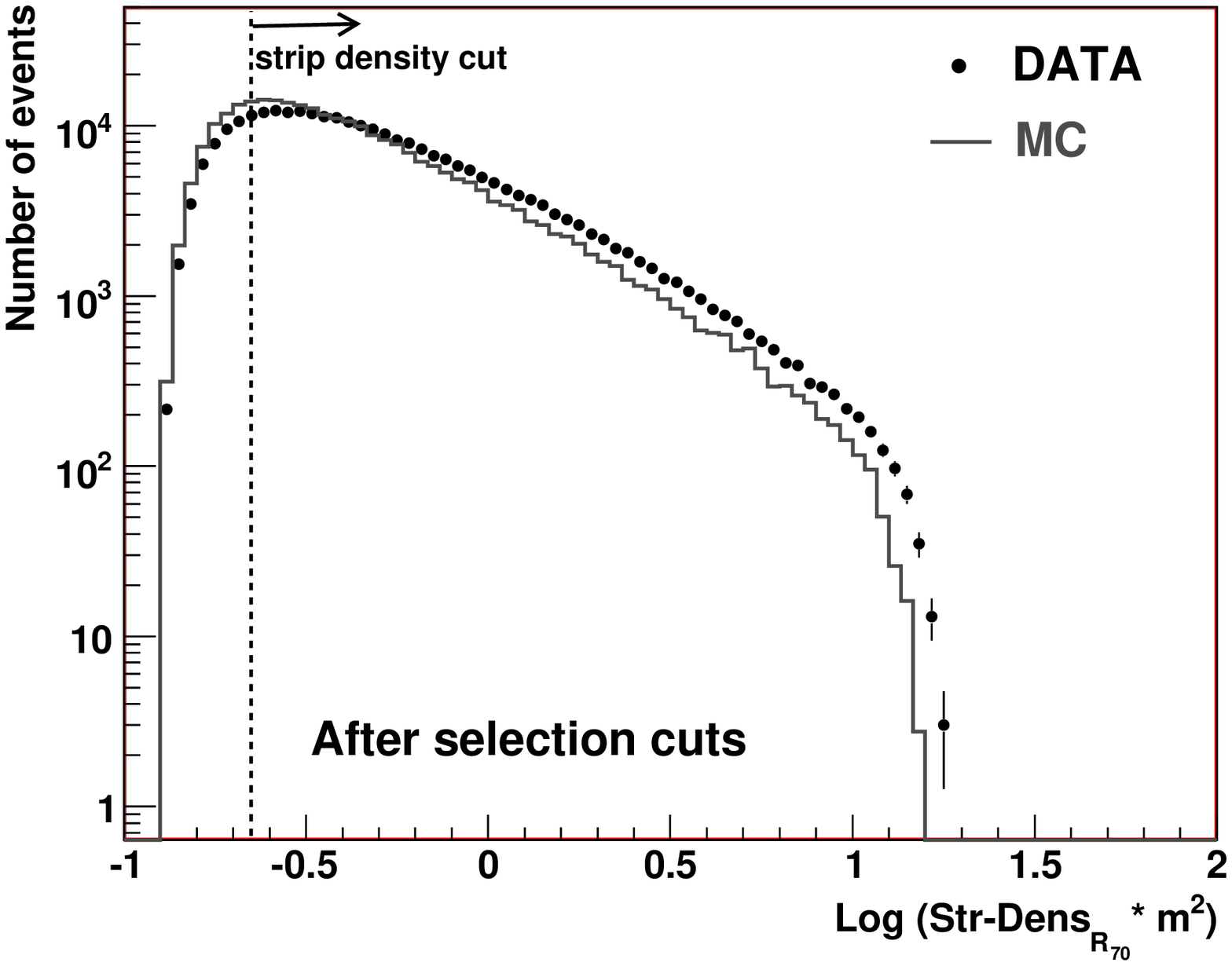} \\
  \end{tabular}
 \end{center}
 \caption{Comparison between simulated and real data before 
          (left) and after (right) the application of selection cuts. 
          The distributions of strip multiplicity are reported in the upper panels. In the lower ones the strip density near 
          the shower core are shown, for events with the reconstructed core position inside the inner 11$\times$8 RPC clusters
          (see text). The simulated sample include also helium initiated showers. As can be seen the agreement is good.
          Real and simulated distributions have been normalized, for comparison, to the same total number of events. }
 \label{DATA_MC}
\end{figure*}

A suitable simulation chain was also set up to check the effects of the different analysis cuts and
to estimate possible systematics.
About  $10^8$ proton-initiated and $2 \times 10^7$ He-initiated showers, with the corresponding power law energy 
spectra between 300$\,$GeV and 3000\,TeV and zenith angle up to 45$^{\circ}$, were produced with the
{\it CORSIKA} code \cite{heck1998}.
In order to have a better evaluation of systematics, we produced independent samples by using
three different hadronic interaction models: {\it QGSJET-I} \cite{kalmykov1997},
{\it QGSJET-II.03} \cite{ostapchenko2006}, and {\it SIBYLL-2.1} \cite{engel1999,fletcher1994}.
As reference values for the energy spectral indexes and the p/He relative normalization 
we used those given in \cite{hoerandel2003} and resulting from a global fit of existing experimental data.
A full simulation of the detector response, based on the GEANT package \cite{geant3}, was performed, including
also the effects of time resolution, trigger logic, electronics noise, etc. 
Simulated data have been produced in the same format used for real ones and they have been analyzed by using 
the same reconstruction code. 
The reliability of the simulation procedure was successfully checked in several ways.
As an example, in Fig.\ref{DATA_MC} there is a comparison between 
real data and the MC simulation, for two of the quantities used in the analysis
As can be seen, the agreement is good both before and after applying the adopted event 
selection cuts (see below). Moreover, as expected, the agreement gets worse if a pure proton composition is 
used for the CR primary flux in the MC sample.

\begin{figure*}
\begin{tabular}{cc}
\includegraphics [width=0.4\textwidth,height=0.4\textwidth]{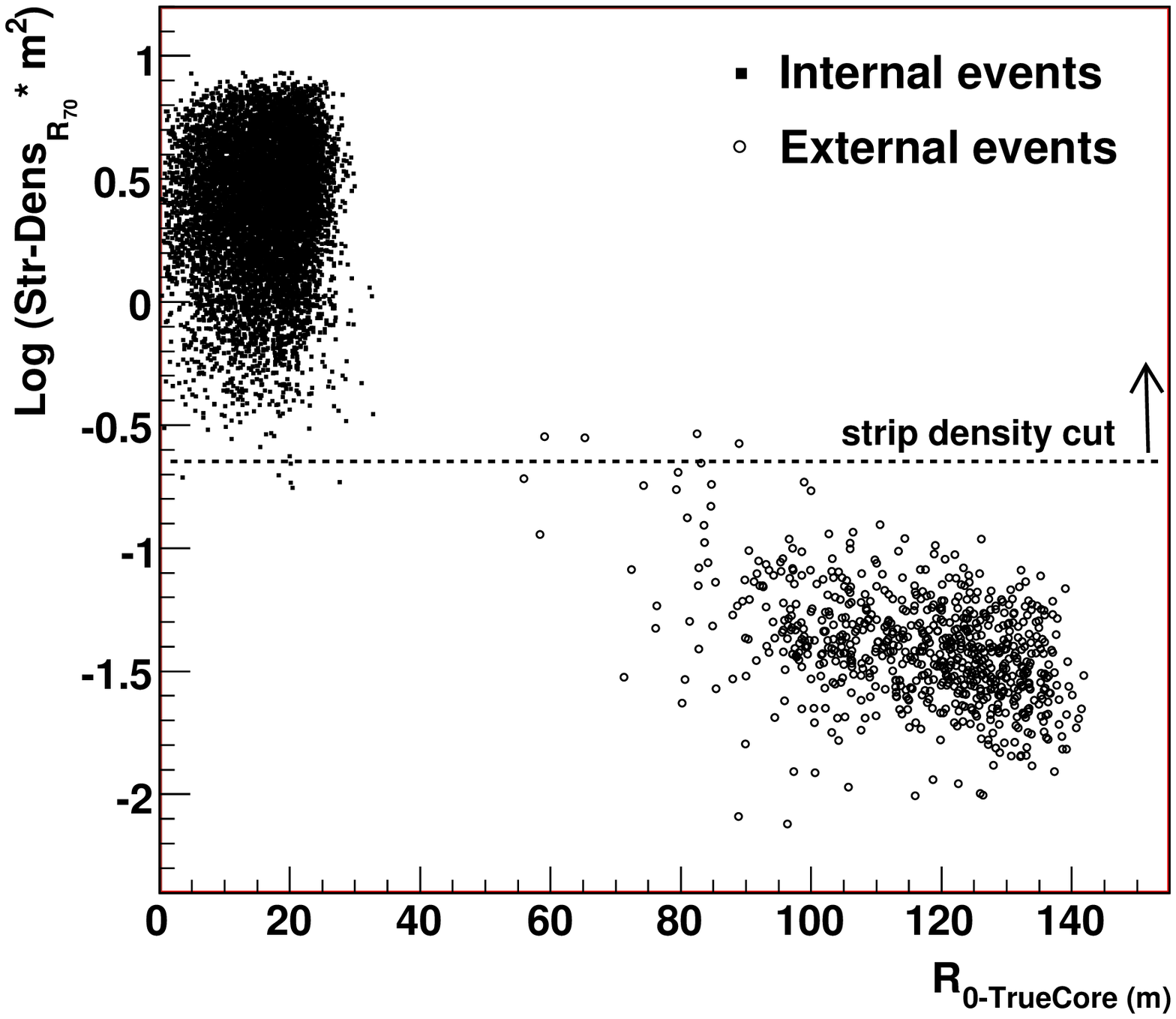} & 
\includegraphics [width=0.4\textwidth,height=0.4\textwidth]{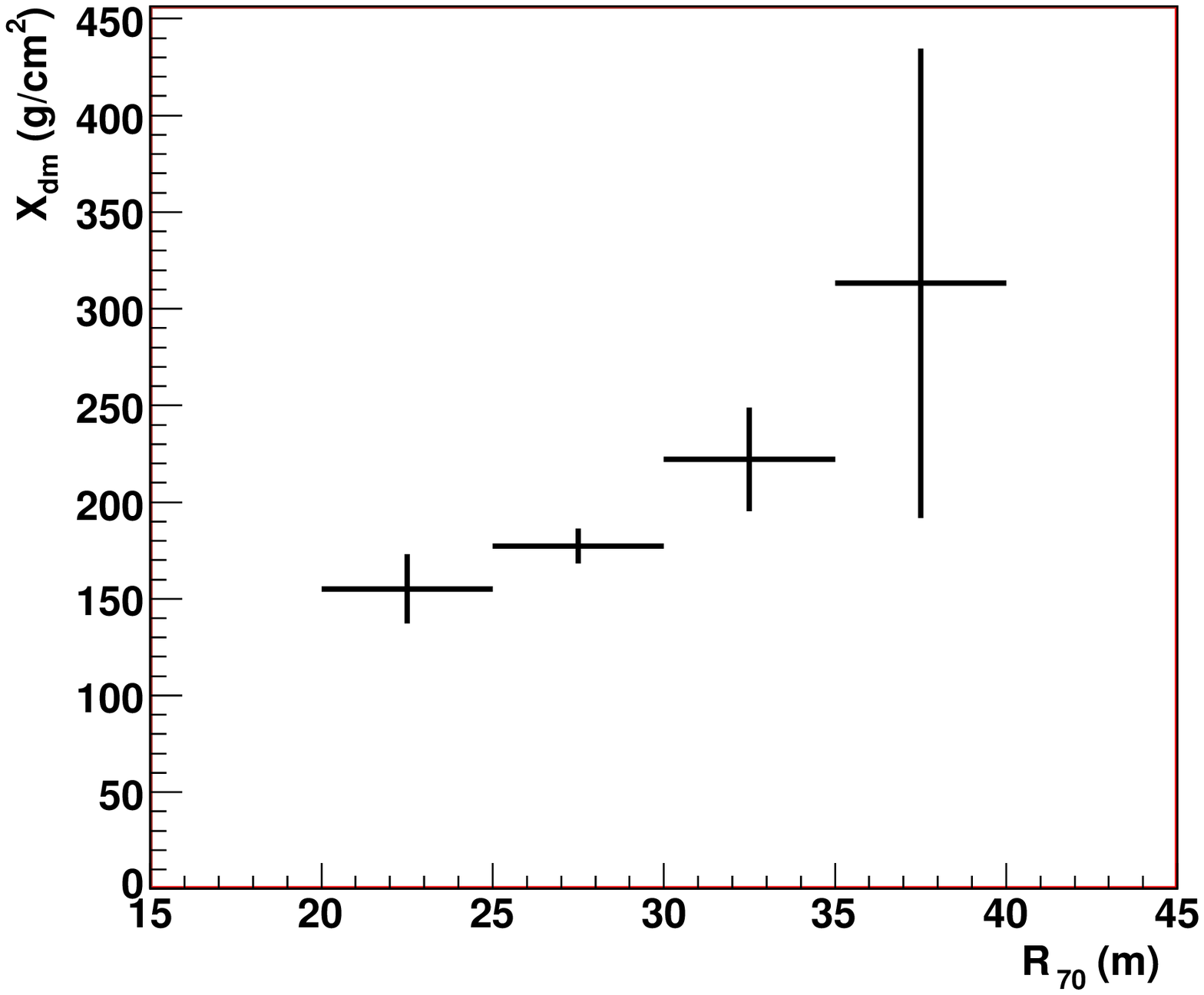} \\
\end{tabular}
\caption{Left panel: Correlation between the strip density near the reconstructed core and distance of the true shower
core from the detector center.
As can be seen, the {\it density cut} (dashed line) is able to efficiently reject misreconstructed events.
Right panel: Correlation between $R_{70}$ and $X_{dm}$. The cut $R_{70} \le 30\,$m allows
the selection of events with small $X_{dm}$ (i.e. deeper shower maximum). 
These plots refer to simulated proton-induced showers processed as real data (see text).}
\label{dens_comp_cuts}
\end{figure*}

\begin{figure*}
 \begin{center}
  \begin{tabular}{cc}
\includegraphics [width=.4\textwidth, height=.4\textwidth]{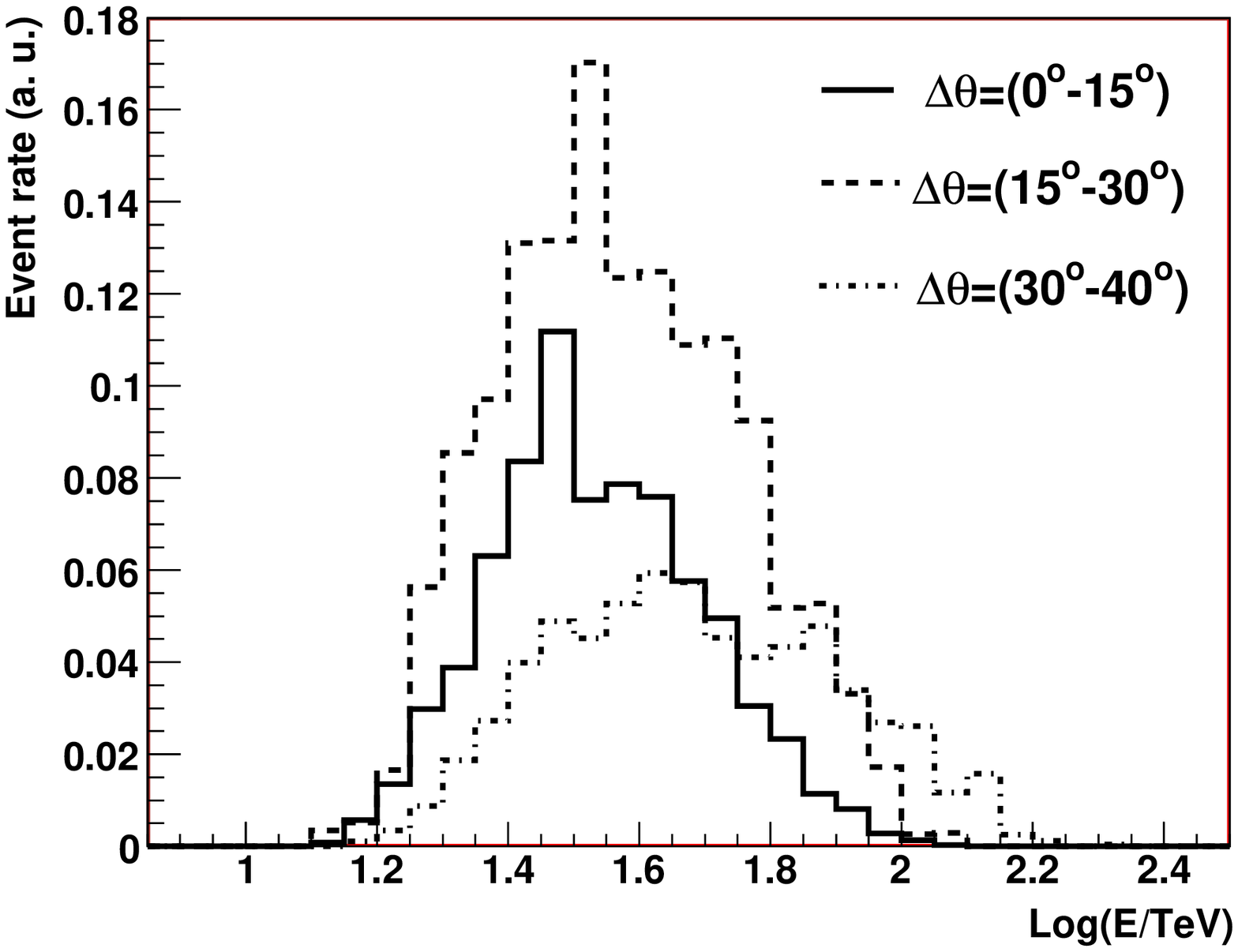}  &
\includegraphics [width=.4\textwidth, height=.4\textwidth]{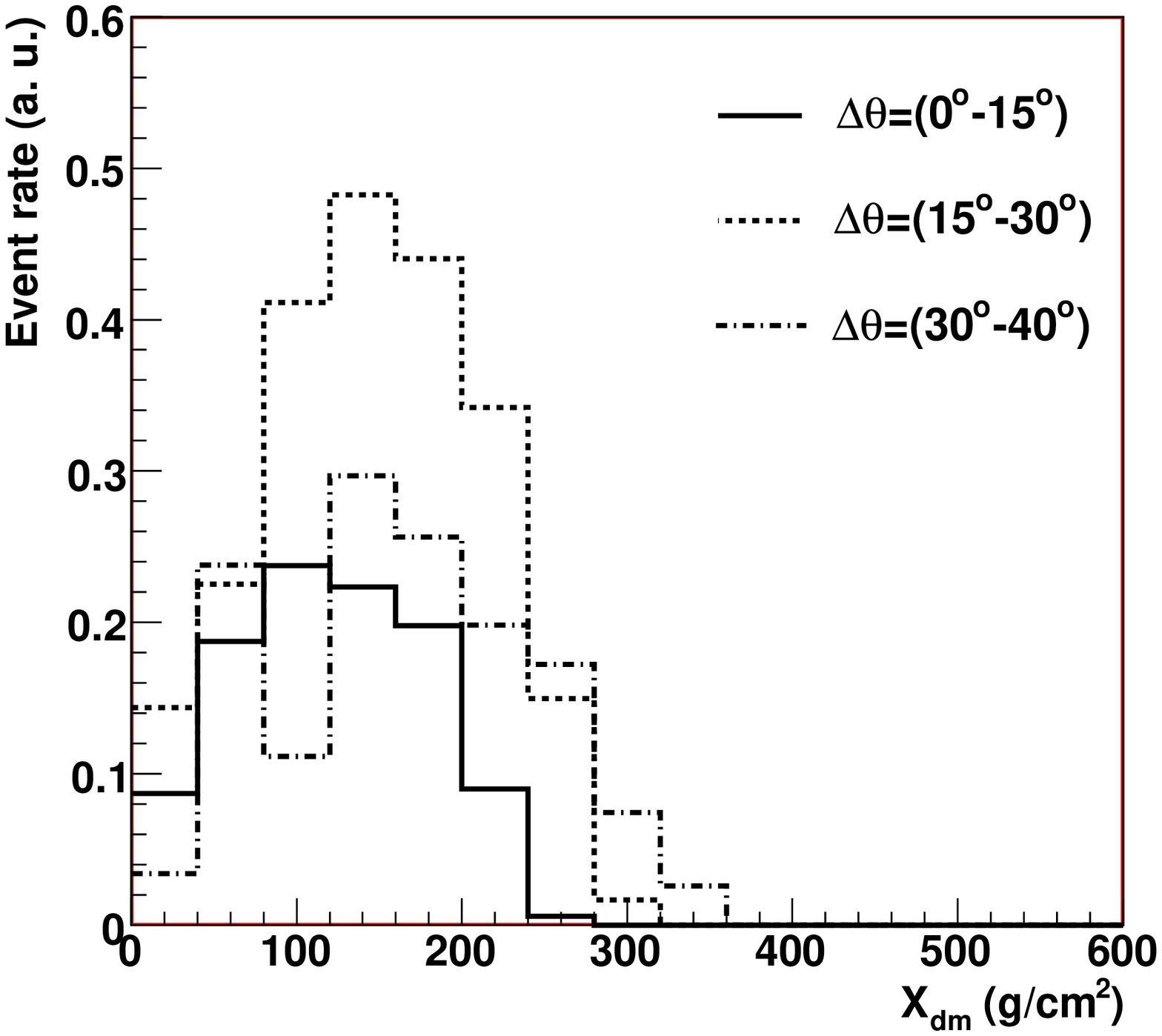}       \\
  \end{tabular}
 \end{center}
 \vskip -0.35cm
 \caption{Energy and $X_{dm}$ distributions (left and right panel, respectively) for proton-initiated MC events, 
          corresponding to one of the strip multiplicity ranges, after the whole event selection procedure has been applied.
          As can be seen the effect of selection cuts does not introduce any substantial zenith-dependent bias.}
\label{Xmax_check}
\end{figure*}

In the analysis of real data, after a first selection based on the quality of the reconstruction procedure, 
a further rejection of {\it external} showers was performed with several additional cuts discussed in the following.
The reconstructed core position, $P_{rc}$, was required to be 
in a fiducial area given by the inner $11 \times 8$ RPC clusters (corresponding to a total surface 
of about $64 \times 64 \,$m$^2$).
This cut reduced the initial data set to about $45 \%$ of the events with reconstructed
$\theta \le 40^\circ$ and $N_{strip} \ge 500$.
The quantity $R_{70}$ was then introduced as the radius of the smallest  circle 
(lying on the detector plane and centered in $P_{rc}$) containing $70\%$ of the fired strips. 
It was then required that the distance of $P_{rc}$ from the detector center plus 
$R_{70}$ should be less than $50 \,$m. The aim of this cut was to select showers largely contained inside the detector 
area (thus very well reconstructed). As a result, the data sample was reduced to $\sim 20 \%$ 
of the initial one.
 One further cut required the minimum average fired strip density, within a distance
$R_{70}$ from the reconstructed core, to be $0.2\,$strips/m$^2$ in the shower plane.
This {\it density cut} allowed the efficient rejection of {\it external} events, 
as shown in Fig. \ref{dens_comp_cuts}, left panel.
The same purpose motivated the last selection cut (a {\it compactness cut}), requiring the 
$R_{70}$ radius to be at most $30 \,$m.
Monte Carlo simulations showed that this {\it compactness cut} is also related 
to the shower development stage, allowing the constraint of the value of $X_{dm}$ (Fig. \ref{dens_comp_cuts}, right panel) 
and the selection of showers with their maximum lying deep in the atmosphere 
(i.e. the sampling of the exponential tail of the $X_{max}$ distribution).
As already pointed out in Sec.\ref{sec:meatechnique}, this is an important point for the extraction 
of cross section data from the $sec \theta$ distribution.
The last two cuts finally selected $12 \%$ of the events 
reconstructed with $\theta \le 40^\circ$ and $N_{strip} \ge 500$ in the initial data sample.
The fractions of events surviving each analysis cut were checked to be consistent with the corresponding quantities 
for MC data. Indeed, once the contribution of CR primaries heavier than protons were considered,
all the values were in agreement and independent of the adopted hadronic interaction model.

The selected data sample was split into five different bins of strip multiplicity $\Delta N_{strip}$ 
(see Tab.\ref{tab:enekappa}), starting from the threshold value of $500$ fired strips on the 
whole central detector (out of the total 124800), in the trigger time window of $420 \,$ns.
As shown in Tab.\ref{tab:enekappa}, each strip multiplicity bin corresponds 
to a different primary energy interval.
In particular the averages of the Log($E$) distributions are reported 
(after the application of all the selection criteria) together with their root mean square values.
The energy values $E$ assigned in this way were checked to be equivalently given by the median energies $E_{50}$ 
of the corresponding event samples, and to be independent of the hadronic interaction model used in the simulation.

Simulations also show that, after applying all the analysis cuts, the contamination 
of {\it external} events misreconstructed as internal is less than $ 1 \%$ for all the energy bins.

In order to avoid any bias in the cross section measurement,
a check has been made that the event selection was independent of the zenith angle. 
As an example, the distributions of primary energies and $X_{dm}$ of simulated events surviving the analysis selection 
criteria are shown in Fig.\ref{Xmax_check}, for different zenith angle intervals.
The distributions are independent of the zenith angle up to about 40$^\circ$, thus showing 
that the experimental sensitivity is not compromised by shower-to-shower fluctuations \cite{alvarez2002,alvarez2004}.

\section{Analysis results and discussion}
\label{sec:discussion}

\begin{figure*}
\begin{center}
\noindent
  \begin{tabular}{ccc}
\includegraphics [width=0.33\textwidth]{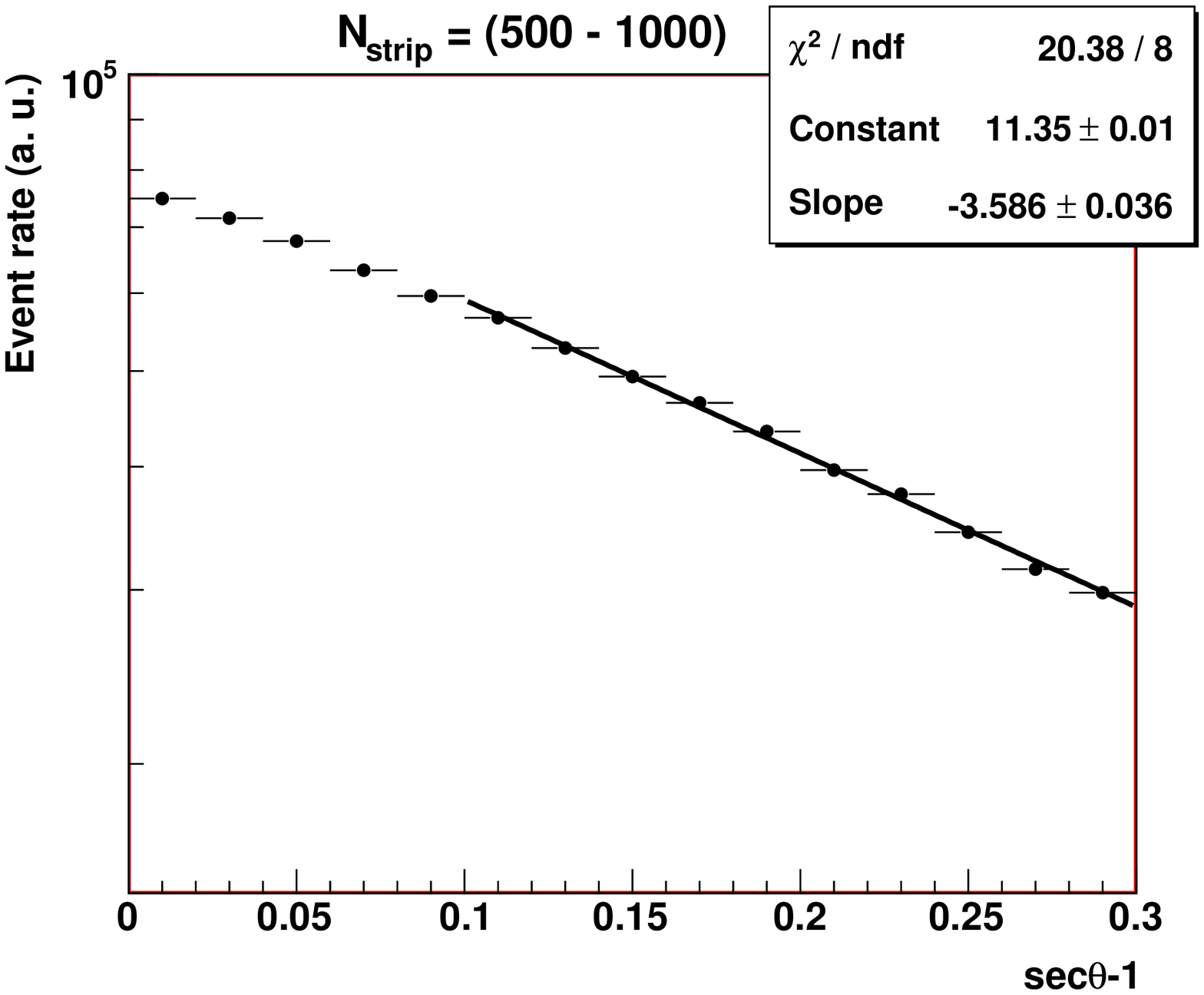} & 
\includegraphics [width=0.33\textwidth]{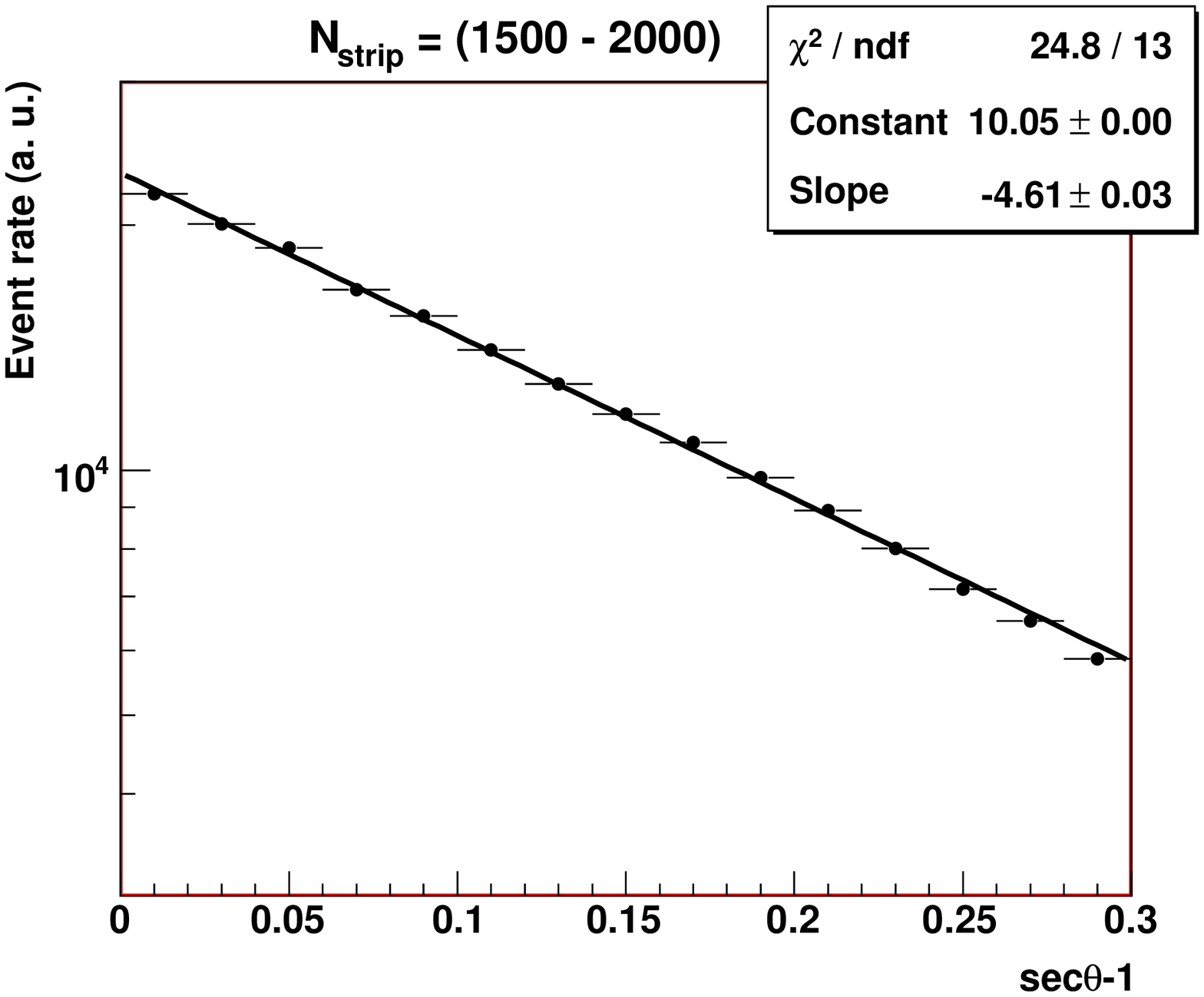} &
\includegraphics [width=0.33\textwidth]{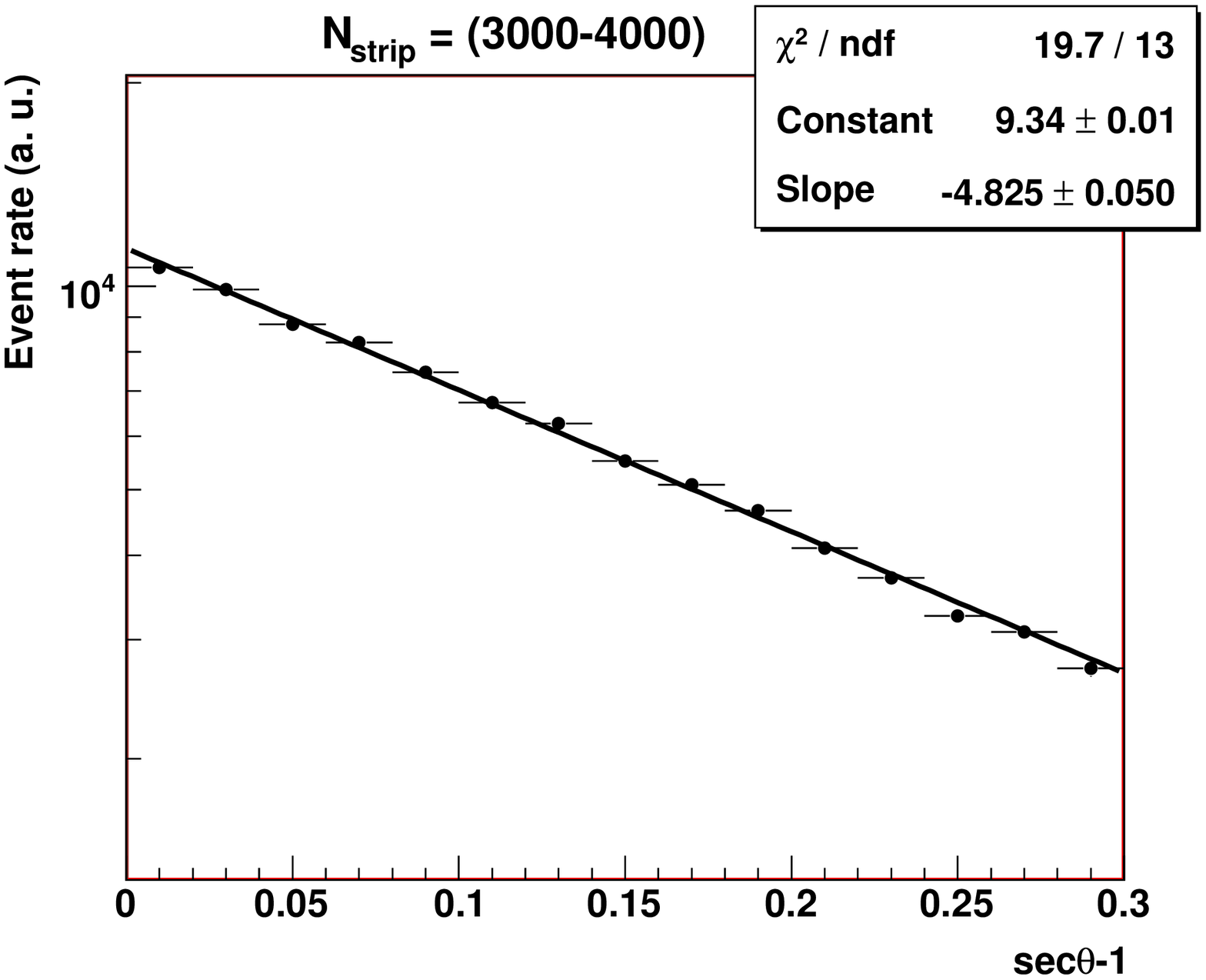} \\ 
  \end{tabular}
  \begin{tabular} {cc}
\includegraphics [width=0.33\textwidth]{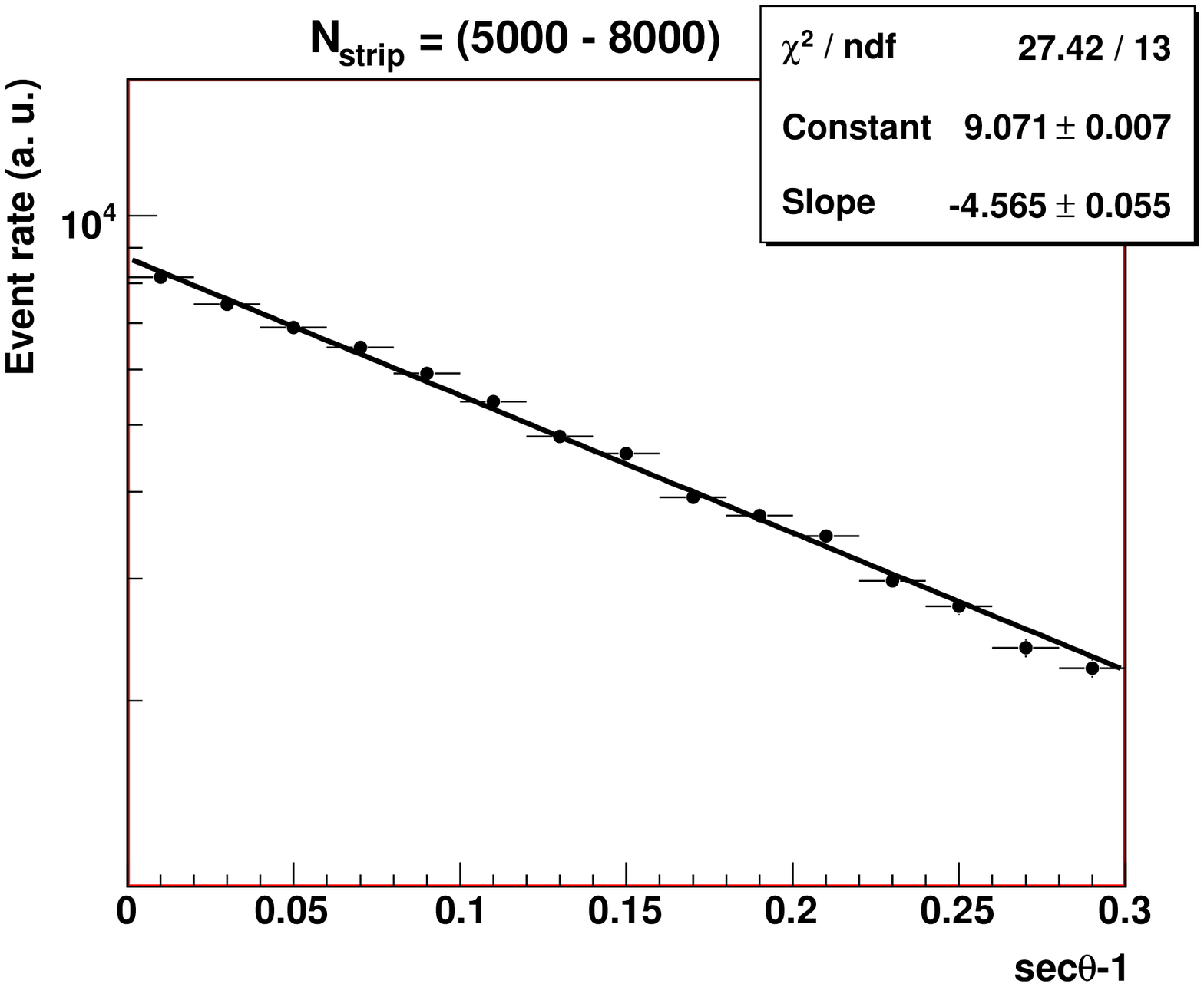} &
\includegraphics [width=0.33\textwidth]{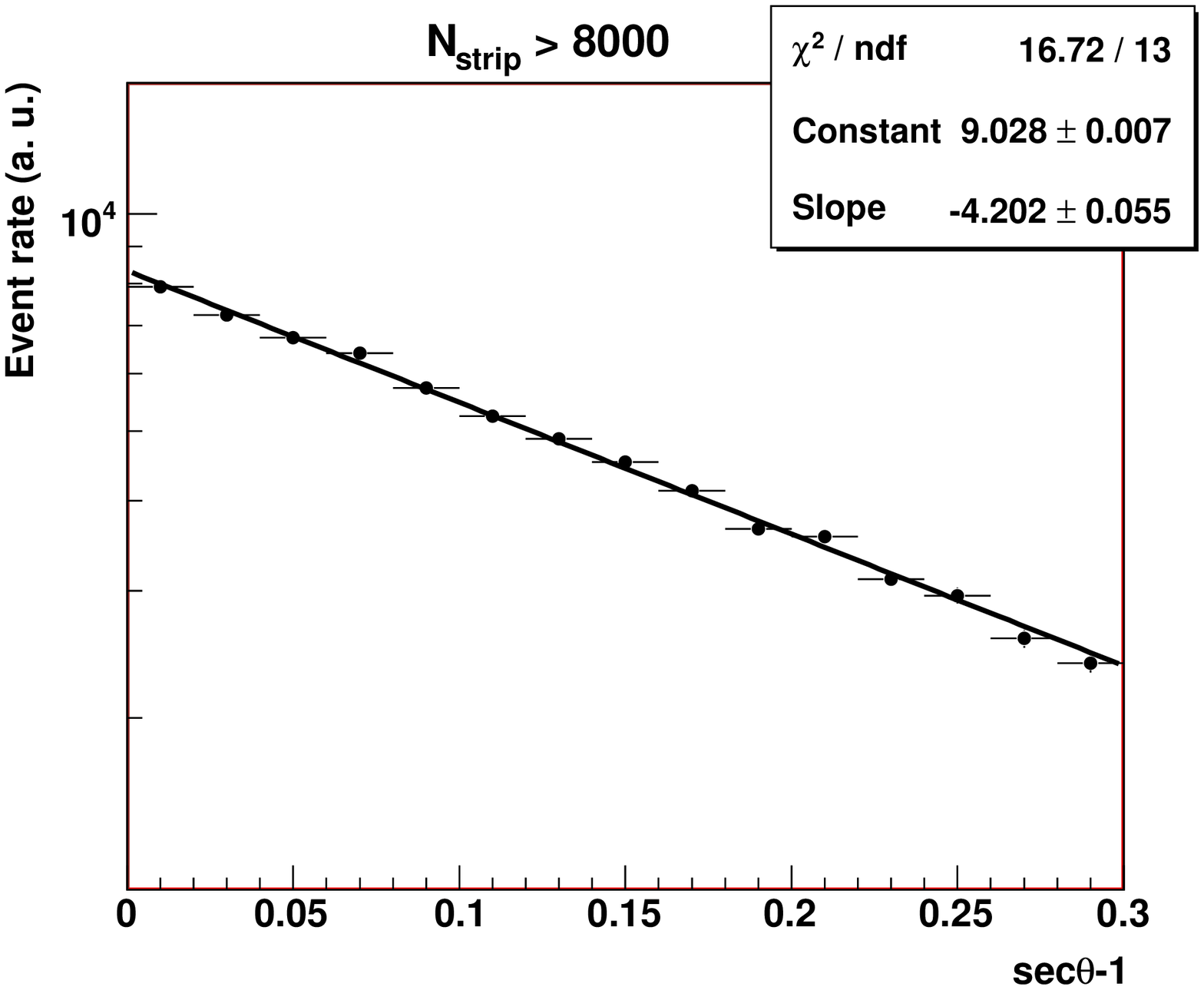} \\
  \end{tabular}
\end{center}
\caption{Experimental $sec\theta$ distributions for the five strip multiplicity ranges, 
 after correction for the geometrical acceptance in each angular bin.}
\label{data_secth}
\end{figure*}

The result of the application of the whole selection procedure to real data is given in Fig.\ref{data_secth}, 
where the experimental $sec \theta$ distributions for the five strip multiplicity ranges are shown,
after correction for the geometrical acceptance of each angular bin.
They clearly show the expected exponential behaviour, this being a further check that the detector capabilities and the 
adopted analysis cuts brought to a proper selection of events for the cross section measurement.
A slight deviation is present, for the lowest energy sample only, at small $sec \theta$ values 
(therefore not included in the fit).  
This is interpreted as due to the effect 
of shower fluctuations (see Sec.\ref{sec:meatechnique}), the shower maximum being more 
distant from the detector for these events. Indeed, for the lowest energy sample 
$\langle X_{max} \rangle \simeq 390\,$g/cm$^2$, while $ \langle X_{max} \rangle \simeq 450\,$g/cm$^2$ for the other ones.

\begin{table*} 
 \begin{center}
\begin{ruledtabular}
  \begin{tabular}{cccccc} 
   $\Delta N_{strip}$ &   Log($E$/eV)    &         $k_{QGSJET-I}$     &     $k_{QGSJET-II.03}$       &     $k_{SIBYLL-2.1}$       &           $k$   \\
   \hline 
   $ 500 \div 1000 $  &  $12.6 \pm 0.3$  &  $1.98 \pm 0.06 \pm 0.05$  &  $1.84 \pm 0.14 \pm 0.05$    &  $1.87 \pm 0.08 \pm 0.04$  &  $1.93 \pm 0.05 \pm 0.06$  \\
   $1500 \div 2000 $  &  $13.0 \pm 0.2$  &  $1.59 \pm 0.03 \pm 0.04$  &  $1.75 \pm 0.12 \pm 0.04$    &  $1.76 \pm 0.06 \pm 0.04$  &  $1.63 \pm 0.03 \pm 0.08$  \\
   $3000 \div 4000 $  &  $13.3 \pm 0.2$  &  $1.69 \pm 0.05 \pm 0.03$  &  $1.63 \pm 0.13 \pm 0.03$    &  $1.72 \pm 0.05 \pm 0.03$  &  $1.70 \pm 0.03 \pm 0.04$  \\
   $5000 \div 8000 $  &  $13.6 \pm 0.2$  &  $1.74 \pm 0.05 \pm 0.03$  &  $1.97 \pm 0.17 \pm 0.04$    &  $1.91 \pm 0.05 \pm 0.03$  &  $1.84 \pm 0.03 \pm 0.10$  \\
   $      > 8000   $  &  $13.9 \pm 0.3$  &  $2.04 \pm 0.06 \pm 0.05$  &  $2.23 \pm 0.19 \pm 0.05$    &  $2.01 \pm 0.05 \pm 0.05$  &  $2.03 \pm 0.04 \pm 0.10$  \\
  \end{tabular}
\end{ruledtabular}
 \end{center}
 \caption{Strip multiplicity intervals, corresponding proton primary energies, and correction factors 
          $k$ with their statistical and systematic uncertainties (see text).}
 \label{tab:enekappa}
\end{table*}

The angular distributions obtained from the MC simulations are very similar to those 
obtained with real data and the same considerations can be applied.
From these plots, the values of $\Lambda^{MC}$ are extracted by fitting them to an exponential function, 
with the slope parameter given by $\alpha = h^{MC}_0 / \Lambda^{MC}$, where 
$h^{MC}_0$ is the nominal atmospheric vertical depth used in the simulation (see below).
Such values, once divided by the values of $\lambda^{MC}_{p-air}$, give the 
parameter $k$ for the different energies.
In particular, for each of the five selected data samples, the averages of the
corresponding distributions of $\lambda^{MC}_{p-air}$ are taken,
while their root mean square values have been used for the evaluation of 
the systematic errors.

As mentioned before, we produced three independent MC samples with
different hadronic interaction models, namely {\it QGSJET-I}, {\it QGSJET-II.03}, and {\it SIBYLL-2.1},
that are widely used for the simulation of EAS and have been shown to provide a consistent description
of their properties \cite{ostapchenko2008}.
In the 1-100$\,$TeV primary energy range, their predictions for $\sigma_{p-air}$ are quite similar,
while there are larger differences for the estimates of the rate 
of diffraction processes and for the inelasticity of proton air interactions \cite{ostapchenko2008,luna2004}.
This last quantity essentially gives the fraction of energy available for particle production 
\cite{shabelsky1992,gaisser1993,frichter1997} and is particularly important for the 
longitudinal development of the shower in the atmosphere. 
Actually the two models that show the largest differences among them are 
{\it QGSJET-I} and {\it SIBYLL-2.1}, while {\it QGSJET-II.03} gives predictions closer to those of {\it SIBYLL-2.1} 
(see for instance \cite{dova2007}).
In order to have a comprehensive and conservative estimate of the systematic errors, we decided to 
consider all of them in the current analysis.

The factors $k$ evaluated with {\it QGSJET-I}, {\it QGSJET-II.03} and {\it SIBYLL-2.1} are reported 
in Tab.\ref{tab:enekappa}, where both the statistical and the systematic 
errors are also shown. The first ones come from the fit procedure, while the second ones from 
the root mean square values of the corresponding $\lambda^{MC}_{p-air}$ distributions. 
As can be seen, there is a general agreement, with larger differences between {\it QGSJET-I} and the other 
two models, as expected. These results are also consistent with the lower average inelasticity of 
{\it QGSJET-II.03} and {\it SIBYLL-2.1}, with respect to {\it QGSJET-I} in the considered energy range (see below).
By following the just mentioned conservative approach, the results have then 
been combined to get the estimated $k$ factors and the associated statistical and systematic errors 
(last column in Tab.\ref{tab:enekappa}).

As already discussed in Sec.\ref{sec:meatechnique}, the value of $k$ 
results from the contribution of many effects that are difficult to disentangle and depend on
hadronic interactions, primary energy, shower-to-shower fluctuations, 
experiment location and capabilities, detection and analysis technique.
As shown in \cite{bellandi1995,shabelsky1992}, for ground array experiments a first contribution to $k$ is given by:
\begin{equation}
 k_0 \simeq \frac{1}{ 1-  \left< (1-K_{in})^{\gamma -1} \right> }
 \label{eq:kappazero}
\end{equation}
where $\gamma$ is the spectral index of the differential CR proton flux,
$K_{in}$ is the p-air inelasticity, and the symbol $\left < ... \right >$ indicates averaging over events.
At our energies the inelasticity is expected in the range $K_{in} \simeq 0.53 \div 0.60$ (depending on the assumed
interaction model \cite{luna2004,alvarez2002simu}), therefore implying $k_0 \simeq 1.25 \div 1.40 $.
Further contributions to $k$ are expected from a flattening of the $sec \theta$ distribution of selected events,
produced by shower fluctuations and detector effects (see Sec.\ref{sec:meatechnique}), thus making $k$ larger. 
This can be parametrized by a factor $k_{det}$, being $k = k_0 \, k_{det}$ \cite{knurenko,ulrich2009}.

\begin{table*}[t]
 \begin{center}
\begin{ruledtabular}
  \begin{tabular}{cccc} 
 $\Delta N_{strip}$ &      $\eta$             & $\sigma_{p-air} \,$(mb) &  $\sigma_{p-p}\,$(mb) \\
 \hline 
 $ 500 \div 1000 $ & $1.00 \pm 0.04 \pm 0.01$ & $272 \pm 13 \pm  9$     &  $43 \pm 3 \pm 5 $ \\
 $1500 \div 2000 $ & $1.00 \pm 0.03 \pm 0.01$ & $295 \pm 10 \pm 14$     &  $48 \pm 3 \pm 6 $ \\
 $3000 \div 4000 $ & $0.99 \pm 0.04 \pm 0.01$ & $318 \pm 15 \pm  8$     &  $54 \pm 4 \pm 6 $ \\
 $5000 \div 8000 $ & $0.98 \pm 0.04 \pm 0.03$ & $322 \pm 15 \pm 20$     &  $56 \pm 4 \pm 7 $ \\
 $      > 8000   $ & $0.95 \pm 0.04 \pm 0.04$ & $318 \pm 15 \pm 21$     &  $54 \pm 4 \pm 8 $ \\
  \end{tabular}
\end{ruledtabular}
 \end{center}
 \caption{Strip multiplicity intervals, correction factors for the helium contribution $\eta$, 
          and proton-air cross sections thus obtained.
          The values of the resulting total proton-proton cross sections estimates are also reported.
          The errors given first come from statistics, the second ones from the systematics.}
          \label{sigmas}
\end{table*}

As can be seen in Tab.\ref{tab:enekappa}, in our case $k \simeq 1.6 \div 1.8$ (then we obtain $k_{det} \simeq 1.15 \div 1.45$ ), 
apart from the values obtained at the boundaries of the covered energy region.
The larger value of $k$ for the lowest energy bin is due to the smaller $\langle X_{max} \rangle$ value, 
that produces a larger effect of shower-to-shower fluctuations (see Sec.\ref{sec:meatechnique}).
An explanation for the relatively high value of $k$ in the highest energy bin is 
given by the onset of saturation of the strip digital information used in the analysis. 
This makes wider the energy interval actually contributing to the considered
multiplicity bin, therefore implying a larger effect of fluctuations, mainly in terms of $X_{rise}$, 
with a consequent loss of sensitivity.
These two effects practically define the energy region in which the current analysis can be performed.
Important improvements will be achieved with the use of the analog RPC readout (see Sec.\ref{sec:argo}).


A check was also made to verify that the detector angular resolution (see Sec.\ref{sec:argo}) is sufficiently good 
in order not to introduce any bias in the cross section measurement. Indeed the slopes $\alpha$ 
obtained by using the true zenith angles $\theta$ or the reconstructed ones were in complete agreement, 
within the quoted errors, for all the five strip multiplicity MC samples.

In the simulations the vertical profile of Linsley's {\it standard atmosphere} was used \cite{heck1998}
with the detection level set at the altitude of 4300 m a.s.l., corresponding to $h^{MC}_0 = 606.7 \,$g/cm$^2$. 
A small source of uncertainty in the p-air cross section measurement (see Eq.\ref{eq:sectheta}) 
might be given by a systematic difference between $h^{MC}_0$ and $h_0$ together with its variations with 
time due to the change of atmospheric conditions.
From the pressure data continuously taken at Yangbajing, we evaluated
$h^{MC}_0/h_0 = 0.988 \pm 0.007$, resulting in an impact on the cross section analysis at the level of $1 \%$,
which has then been taken into account for the cross section data presented here.

As outlined in Sec.\ref{sec:meatechnique}, the measured $\Lambda^{exp}_{CR-air}$ value together 
with the $k$ factor determined from the simulation, 
directly gives the experimental interaction length $\lambda^{exp}_{CR-air}$ and consequently the production 
cross section ${\sigma}_{CR-air}$. At this stage, the cross section has still to be corrected for the contribution of 
CR primaries heavier than protons.
This correction has been estimated by evaluating the effect 
of the introduction of helium primaries in the simulated data
on the shape of the $sec \theta$ distribution.
Corrections for other primaries (i.e. CNO group, Fe, etc.) are negligible.
Both the different absolute values and the energy dependences of proton and helium fluxes were 
considered, by taking as a reference the fits to the experimental data given in \cite{hoerandel2003}.
As expected, the simulations showed a slight steepening of the $sec \theta$ distribution at the
highest strip multiplicity intervals, thus changing the values of the
corresponding cross section estimates.
By applying this procedure we got the correction factors $\eta$ to be applied to ${\sigma}_{CR-air}$ 
in order to obtain $\sigma_{p-air}$ (see Tab.\ref{sigmas}).
It has also been checked that the effect of primary particles heavier than protons 
on the shape of the $sec \theta$ distribution, is limited to few per cent not only due to the CR beam composition itself, 
but also because the analysis cuts actually select a proton-enriched sample. 
Indeed, the initial fraction of He-induced showers is further reduced by about a factor 1.9, 
through the analysis flow, with respect to proton-induced events.

The effect of the uncertainty of the primary CR composition has been estimated by applying 
the same procedure starting from other primary flux measurements, 
namely those from the JACEE\cite{asakimori1998} and RUNJOB\cite{apanasenko2001} experiments. 
We found small differences (below $4 \%$) which were used for the evaluation of the 
systematic uncertainty on the factor $\eta$ (see Tab.\ref{sigmas}).

\begin{figure*}
 \begin{center}
  \includegraphics [width=0.9\textwidth,height=0.65\textwidth]{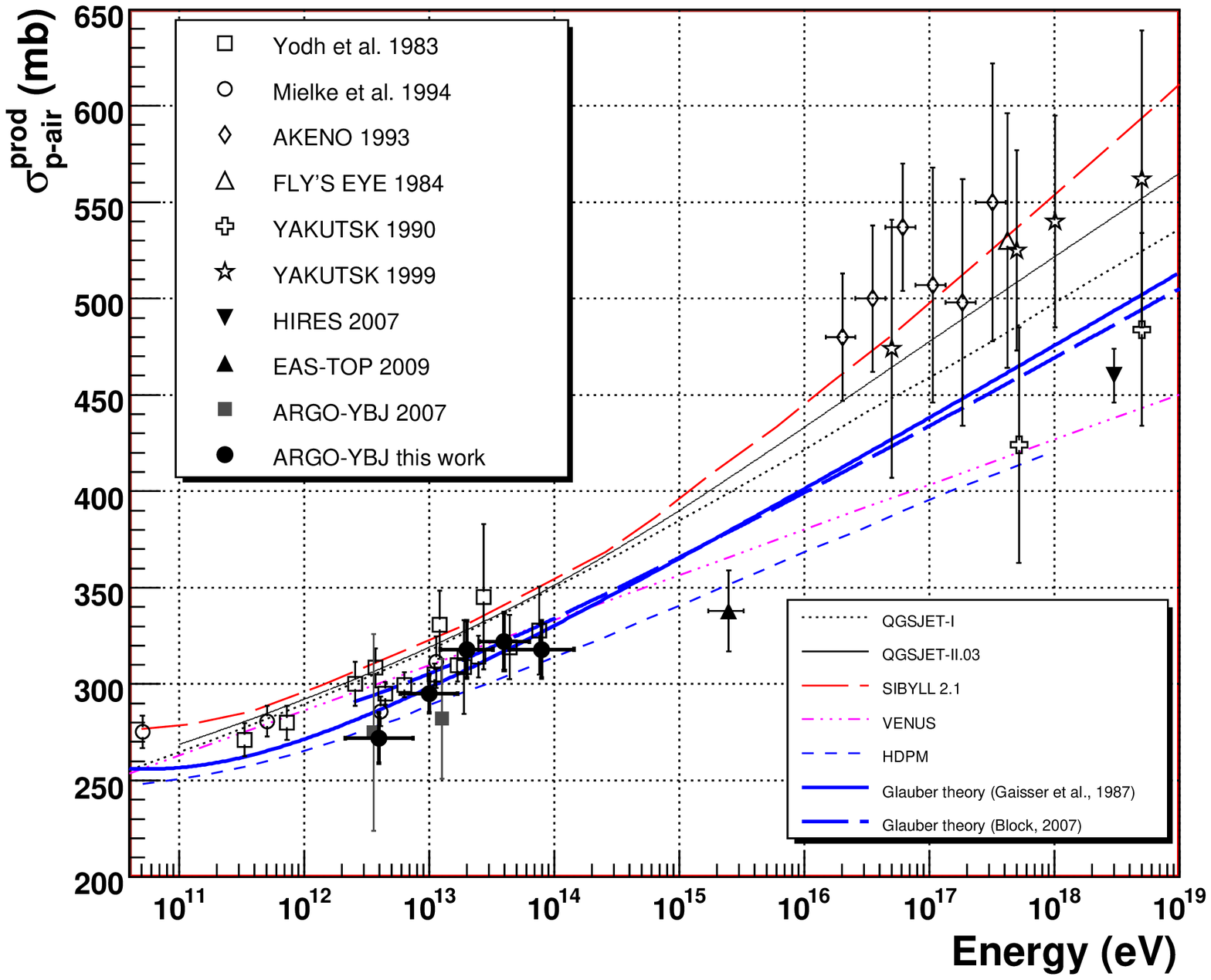}
 \end{center}
 \caption{Proton-air {\it production} cross section measured by ARGO-YBJ and by other CR experiments
          \protect\cite{yodh1983,gaisser1987,mielke1994,honda1993,knurenko,aglietta2009,baltrusaitis1984,belov2007,demitri2007} 
           together with the values given by several hadronic interaction models \protect\cite{heck1998,knapp2003,ulrich2007dev}.
           Also shown are the predictions of two different calculations based on Glauber theory \cite{gaisser1987,block2007} 
           applied to accelerator data as  parametrized in \cite{blockhalzen2005}.
           In this plot ARGO-YBJ points have been already corrected for the effects of CR primaries heavier than protons.           
         }
\label{fig:pair}
\end{figure*}

The resulting proton-air production cross sections, $\sigma_{p-air}$, are summarized in Tab.\ref{sigmas}, 
where both statistical and systematic errors are reported.
These results are consistent with a previous ARGO-YBJ measurement done without using the strip information \cite{demitri2007},
and also with a preliminary estimate obtained from the analysis of a data set taken with only 42 RPC clusters 
during the detector installation \cite{isvhecri2006,ecrs2006}.

The measured p-air {\it production} cross section is reported, in Fig.\ref{fig:pair}, as a function of the primary proton energy. 
The results found by other experiments and the expectations given by some hadronic interaction models are also 
shown \cite{heck1998,knapp2003,ulrich2007dev}.
Even if data are in general agreement, it can be noticed some discrepancy between the various experimental approaches
that might be due to the different method systematics (see Sec.\ref{sec:intro}).
Furthermore, new results, namely those from ARGO-YBJ, EAS-TOP\cite{aglietta2009} and HiRes\cite{belov2007}, systematically give 
cross section values that are slightly lower with respect to the more recent and comprehensive hadronic 
interaction models actually used in these analyses.
The indication for lower cross section (and/or inelasticity) values is also consistent with what found from the analysis 
of several other EAS observables \cite{hoerandel2003xsection,erlykin2007,abdallah2007}.
Indeed, as previously pointed out, the inelasticity distribution in p-air interactions plays an important role in the 
longitudinal development of atmospheric showers. In EAS-based cross section measurements these effects are taken into account 
by the $k$ factors. Anyway residual systematic differences between data and expectations at the level of few per cent
might result from the modeling and the effect of very low inelasticity collisions (i.e. $K_{in} \lesssim 0.05$) that cannot 
actually be observed, being due to diffractive interactions that weakly (or do not) influence the EAS development 
\cite{luna2004,knapp2003,ulrich2007dev}, while are included in the cross section values given by the hadronic models 
in Fig.\ref{fig:pair} \cite{heck1998,alvarez2002simu}.
This kind of effects might be different in the case of cross section evaluations based on 
measurements of the single hadrons flux at ground \cite{yodh1983,gaisser1987,mielke1994}.
The low energy threshold of ARGO-YBJ (with respect to other EAS experiments) allows a direct comparison 
with the values given by this technique, showing a good agreement. This is particularly important since
the systematics of the two measurement techniques are completely different. 
The agreement also extends to the predictions of different calculations based on the Glauber theory \cite{glauber1970},
applied to the measurements made at particle accelerators. 
As an example, the results of two of them \cite{gaisser1987,block2007} are actually shown in Fig.\ref{fig:pair}, 
starting from the accelerator data analysis performed in \cite{blockhalzen2005}, where 
an asymptotic $ln^2(s)$ dependence of hadronic cross sections is shown to be favoured.

\begin{figure*}
  \begin{center}
\includegraphics [width=0.9\textwidth,height=0.65\textwidth]{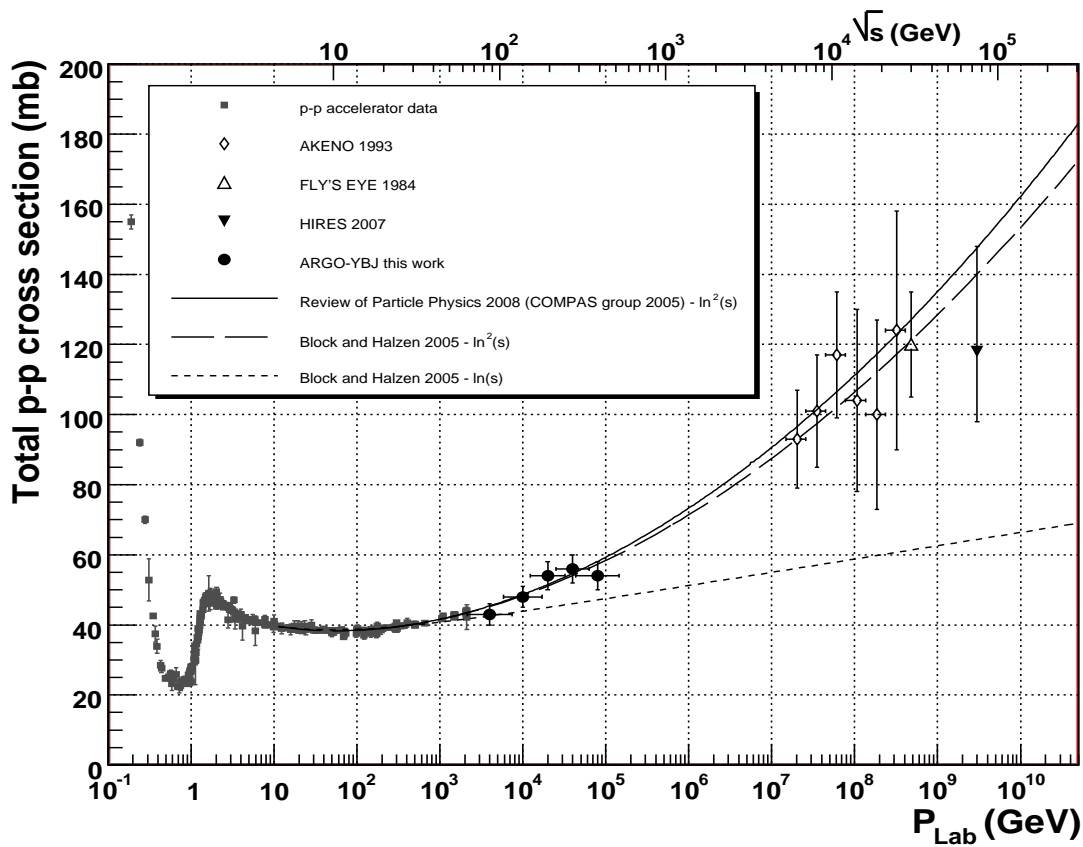}
  \end{center}
  \caption{Total p-p cross section, $\sigma_{p-p}$, obtained by ARGO-YBJ starting from $\sigma_{p-air}$
           together with the same quantity published by other CR experiments \cite{honda1993,baltrusaitis1984,belov2007}.
           The values of the total p-p cross section measured at accelerator experiments \cite{pdg2008} are also reported. 
           The lines result from a global analysis of accelerator data \cite{pdg2008,blockhalzen2005,cudell2002}, 
           assuming an asymptotic behaviour of hadronic cross sections proportional to $ln^2(s)$ or $ln(s)$, as reported in the plot.}
  \label{sigma_pp}
\end{figure*}

As outlined in Sec.\ref{sec:intro}, Glauber theory can also be used to estimate 
the total proton-proton cross section $\sigma_{p-p}$, at energies not covered by accelerator data,
by starting from the p-air {\it production} cross section $\sigma_{p-air}$ as measured by CR experiments.
These calculations were discussed in several papers
\cite{block2006,engel1998,gaisser1987,durand1990,kopeliovich1989,bellandi1995,wibig1998,block2007}.
As pointed out by many authors \cite{block2006}, an essential step in the conversion is the knowledge of the
dependence of the slope, $B$, of the forward scattering amplitude for elastic p-p collisions on 
the center of mass energy $\sqrt{s}$, or on $\sigma_{p-p}$ itself. 
For ARGO-YBJ data, we applied the conversion given in \cite{gaisser1987}, that uses the somewhat model-independent Chou-Yang 
prescription \cite{chouyang1983} between $B$ and the total cross section $\sigma_{p-p}$, 
this being also consistent with available measurements on $B(s)$.
The results were compared with what given by other models \cite{durand1990,kopeliovich1989,bellandi1995,wibig1998,block2007},
the larger differences being with \cite{bellandi1995} and \cite{wibig1998}, that foresee
slightly higher or lower $\sigma_{p-p}$ values, respectively.
Anyway in all the cases the differences, in our energy range, are below $10\%$ and they have been added as 
a further contribution to the systematic error on the resulting $\sigma_{p-p}$.
The results are summarized in Tab.\ref{sigmas} and in Fig.\ref{sigma_pp}.
Also shown in the figure are the measurements performed at accelerators \cite{pdg2008} and the $\sigma_{p-p}$ values 
published by other CR experiments \cite{honda1993,baltrusaitis1984,belov2007} starting from $\sigma_{p-air}$.
As can be seen, the ARGO-YBJ data lie in an energy region not yet reached by p-p colliders (and 
still unexplored by p-$\bar{\rm p}$ experiments \cite{geich-gimbel1989,avila2002,abe1994}). 
Our result is in agreement with the general trend of accelerator data showing the rise of the
cross section with energy. 
In particular, it favours the asymptotic $ln^2(s)$ increase of total hadronic
cross sections as obtained in \cite{blockhalzen2005} from a global analysis of accelerator data.

\section{Conclusions}
\label{sec:conclus}

The study of CR-induced showers provides a unique opportunity to explore hadron 
interactions in energy ranges and phase space regions currently not accessible at particle colliders.
In this paper, the cosmic ray flux attenuation for different atmospheric depths (i.e. zenith angles) has been exploited
in order to measure the proton-air cross section in a laboratory energy range between approximately 1$\,$TeV and 100$\,$TeV,
with the ARGO-YBJ experiment. 
The information provided by a highly segmented full coverage detector, allowed the introduction of selection cuts 
able to both define the energy scale and constrain the shower development stage.
The analysis results have been also used to estimate the total proton-proton cross section at center of mass energies
between 70 and 500$\,$GeV.

The analysis was applied to a data sample of about $6.5 \times 10^8$ events.
A full simulation of the primary interaction, the shower development in the atmosphere, the particle detection at ground 
and the complete analysis procedure has also been setup and used in order to make consistency checks and evaluate the method 
systematics.
The effects of shower fluctuations and the contribution of heavier primaries have been also considered. 
Moreover the use, in the simulations, of three different hadronic interaction models allowed a comprehensive and conservative
estimate of the associated systematic uncertainties.

The results for the proton-air {\it production} cross section have been compared with previous measurements and model predictions.
A good agreement is present with other measurements made in the same energy region with completely different techniques, and also
with theoretical predictions based on particle accelerator data.
Some systematic differences between more recent results and the values given by commonly adopted hadronic interaction models are 
present at the level of few per cent, the measured cross sections suggesting lower values 
and/or deeper shower development in the atmosphere.

The total proton-proton cross section $\sigma_{p-p}$ has then been inferred from the measured proton-air 
{\it production} cross section $\sigma_{p-air}$ by using the Glauber theory. 
The resulting data lie in an energy region not yet reached by p-p colliders and so far unexplored by p-$\bar{\rm p}$ experiments. 
Our result is consistent with the general trend of experimental data, favouring an asymptotic $ln^2(s)$ rise of the cross section.

Further improvements in the analysis are expected from the use of the detailed information on the shower front 
(curvature, rise time, time width, etc.) that ARGO-YBJ is able to record with very high precision, and by the 
implementation of the analog RPC readout that will allow extending these studies 
to collisions with center-of-mass energies up to the TeV region.

\begin{acknowledgements}
Useful discussions with R.~Engel, P.~Lipari, D.~Heck, S.~Ostapchenko, T.~Pierog and R.~Ulrich are gratefully acknowledged.
This work is supported in China by NSFC (No. 10120130794), the Chinese Ministry of Science and Technology, 
the Chinese Academy of Sciences, the Key Laboratory of Particle Astrophysics, CAS, and in Italy by the Istituto 
Nazionale di Fisica Nucleare (INFN) and the Ministero dell'Istruzione, dell'Universit\`a e della Ricerca (MIUR).
\end{acknowledgements}

 

\end{document}